\documentclass[fleqn,usenatbib]{mnras}
\usepackage{longtable}
\usepackage{newtxtext,newtxmath}
\usepackage[T1]{fontenc}
\DeclareRobustCommand{\VAN}[3]{#2}
\let\VANthebibliography\thebibliography
\def\thebibliography{\DeclareRobustCommand{\VAN}[3]{##3}\VANthebibliography}

\usepackage{graphicx}	% Including figure files
\usepackage{amsmath}	% Advanced maths commands
\newcommand{\unp}{t.w.}
\newcommand{\tw}{{\bf t.w.}}
\newcommand{\spl}{{\bf (s.l.)}}
\newcommand{\splc}{{\bf (s.l.:)}}
\newcommand{\vi}[2]{\ensuremath{#1} & $\pm$ & #2}
\newcommand{\va}[2]{\multicolumn{3}{c}{\ensuremath{#1}}}
\newcommand{\nulla}{\multicolumn{3}{c}{--}}
\title[]{Magnetism among the oldest white dwarfs in the solar neighbourhood}

\author[Bagnulo \& Landstreet]{
Stefano Bagnulo,$^{1}$\thanks{E-mail: stefano.bagnulo@armagh.ac.uk}
and John D. Landstreet$^{1,2}$
\\
$^{1}$Armagh Observatory \& Planetarium, College Hill, Armagh BT61 9DG, UK\\
$^{2}$Dept. of Physics \& Astronomy, University of Western Ontario, London, Ontario N6A 3K7, Canada\\
}

\date{Accepted 2026 October 30. Received 2026 October 30; in original form 2026 September 29}

\pubyear{2026}

\begin{document}
\label{firstpage}
\pagerange{\pageref{firstpage}--\pageref{lastpage}}
\maketitle

\begin{abstract}
We present a new circular polarisation survey of approximately 100 white dwarfs selected as featureless DC stars and combine our measurements with previous observations. Our compilation contains 205 objects, of which 24 show firm circular polarisation detections, while another four show marginal signals that require confirmation. Four magnetic white dwarfs originally classified as DC show polarisation spectra with significant contributions from spectral lines. Depending on the treatment of these objects and of the marginal detections, the inferred fraction of magnetic objects in the DC selected sample lies between approximately 10 and 14\,\%. Among objects without identifiable line contributions, the circular polarisation never exceeds approximately 2\,\%, substantially below the levels observed in some hotter magnetic white dwarfs with prominent spectral features. This is unlikely a 
 symptom of field decay, since in our sample we find no significant evidence for a dependence of magnetic incidence on cooling age, and magnetic fields are detected in stars with cooling ages exceeding 9\,Gyr. Instead, strong spectral line opacity provides a natural mechanism for producing much larger polarisation amplitudes than the continuum. In the full sample considered here, the incidence of detectable magnetism increases with stellar mass. Most average-mass magnetic stars for which repeated observations are available show no clear polarimetric variability, whereas variability appears more common among the most massive objects.
\end{abstract}


%%%%%%%%%%%%%%%%%%%%%%%%%%%%%%%%%%%%%%%%%%%%%%%%%%

%%%%%%%%%%%%%%%%% BODY OF PAPER %%%%%%%%%%%%%%%%%%

\section{Introduction}
Magnetic fields are detected in a substantial fraction of white dwarfs, with strengths spanning from a few kG to several hundred MG. Despite decades of observational and theoretical work, their origin and evolution remain uncertain. Proposed formation channels include the survival of fields generated during earlier evolutionary stages, as well as field generation associated with binary interaction or merger events \citep[see][and references therein]{Feretal20}, and it is likely that more than one mechanism contributes to the observed population \citep{ValFab99,BagLan22,Mosetal25}. 

Although substantial progress has been made in establishing observational constraints on magnetism in WDs, these constraints are not uniform across the white-dwarf population. Volume-limited polarimetric surveys have provided robust measurements of the incidence of magnetism in nearby white dwarfs \citep{Kawetal07,BagLan21,BagLan22,Mosetal25}. However, determining the magnetic incidence as a function of cooling age and mass becomes increasingly difficult for older and cooler stars, because the spectral lines through which magnetic fields are most readily detected and measured become progressively weaker, or disappear altogether. This introduces an important observational bias in the regime where magnetic incidence is expected to be especially informative. In addition, surveys differ in observational technique, detection threshold, and sample selection, making it difficult to compare magnetic incidence across spectral classes  in a fully homogeneous way \citep{BagLan18,BagLan21}.

Featureless, or DC, white dwarfs represent a particularly important and challenging target population in this context. These stars show no lines in their optical spectra. In such objects, magnetic fields cannot be identified through spectral-line splitting or line polarisation, but only through continuum circular polarisation. Circular polarimetry is therefore the only direct observational tool available to search for magnetism in this class. At the same time, this method is mainly sensitive to relatively strong fields (i.e., at least 5--10\,MG), so that the magnetic incidence among DC WDs remains much less well constrained than in spectral classes with measurable line profiles \citep[see, e.g., Sect.~3 of][]{BagLan20}.

Volume-limited studies have shown that magnetism is uncommon among the youngest, average-mass WDs, but becomes significantly more frequent among older stars \citep{ValFab99,BagLan21}. From the point of view of magnetic-field evolution, DC white dwarfs are therefore especially interesting. They include some of the oldest known stars in the solar neighbourhood, and probing them through continuum circular polarimetry is the only direct way to test whether magnetic fields persist to the oldest observable stages of white dwarf cooling.

Several observational programs have searched for magnetic fields in DC WDs. Early broadband circular-polarisation surveys demonstrated that strong fields can be detected through continuum polarisation, and established the basic feasibility of this approach \citep{Kemetal70,AngLan71-Second,Angetal81}. Later work extended the search for magnetic white dwarfs using circular spectropolarimetry \citep{Putney97}, while more recent investigations have shown that modern broadband polarimeters can efficiently detect magnetic DC stars and significantly enlarge the known sample of such objects \citep{Beretal22,Beretal23,Beretal24}.

Here we present new circular-polarimetric observations of featureless WDs obtained with the FORS2 instrument at the ESO Very Large Telescope. Our dataset includes both filter polarimetry and spectropolarimetry, and we also compile published circular-polarisation measurements of DC WDs from the literature, with the aim of constructing the most complete observational dataset currently available for this class.  

It is important to stress that the DC classification does not correspond to a unique physical category, but rather to a taxonomic class defined by the absence of detectable spectral features. Helium-atmosphere WDs become featureless at significantly higher temperatures than hydrogen-rich stars, while sufficiently strong magnetic fields may broaden, distort, or suppress spectral lines to the point that the optical spectrum appears essentially featureless. As a consequence, the DC class is intrinsically heterogeneous. We also note that some of the stars originally classified as DC may reveal spectral lines in higher resolution or higher S/N ratio observations.

 Our new dataset of about 100 white dwarfs is similar in size to the one obtained by \citet{Beretal22,Beretal23,Beretal24}, who observed about 80 DC white dwarfs with the DiPOL polarimeter attached at the Nordic Optical Telescope. Literature data preceding the work by Berdyugin et al. consist of observations of about 30 DC white dwarfs \citep[e.g.,][ for more details see Sect.~\ref{Sect_Combi}]{AngLan71-Second,LieSto80,Angetal81,Putney97}.

Our primary goal is to provide an updated estimate of the incidence of magnetic fields among cool DC white dwarfs, improving on earlier determinations based on smaller samples, and address a specific evolutionary question: whether magnetic fields persist among the oldest and coolest WDs accessible to observation. The results presented here are not intended to provide, by themselves, a complete constraint set for the physical interpretation of magnetic-field evolution in WDs, but a specific observational constraint on the occurrence of magnetic fields at the cool end of the WD population. In other words, for the purpose of studying the evolution of the magnetic fields in white dwarfs, the data presented in this paper represent an important complement to other observational studies but should not be taken in isolation. Furthermore, it should be emphasised that, with current instrumentation, detectable continuum circular polarisation is generally produced only by relatively strong magnetic fields. Since circular polarisation is sensitive to the line of sight component of the magnetic field, its detectability also depends on the field geometry and viewing angle. We estimate that, for featureless white dwarfs, the detection threshold typically corresponds to a mean surface field strength of approximately 10\,MG.

A further aim of this work is to investigate the variability of the circular polarisation signal in magnetic DC WDs. Previous studies have suggested that some old, strongly magnetic white dwarfs may show little or no polarimetric variability over long timescales \citep{SchNor91,Feretal20}; in fact, \citet{BagLan24} found that non-variability is a ubiquitous feature of stars older than a few Gyr with fields stronger than $\sim 20$\,MG. By combining our observations with literature data, we can test this behaviour for a larger sample of strongly magnetic white dwarfs, most of which are older than a few Gyr. We will also discuss whether some of these objects with constant circular polarisation may be useful as standard stars for circular polarisation. 

\section{New observations}
 The targets of our new observations were mostly selected among the DC white dwarfs of the local 40\,pc volume \citep[we mainly used the spectral classification given by][]{OBretal24}, with particular attention towards the oldest stars in order to check for evidence of ohmic decay. In order to best sample the oldest age bin, some of our oldest targets are beyond the local 40\,pc volume. Targets of our new observations are therefore slightly biased in favour of the oldest stars. 
 
 We obtained 130 polarimetric observations (corresponding to 155 distinct measurements, because in some observations we employed more than one filter or grism) for 97 white dwarfs, for a total of $\sim 56$\,h shutter time, either in imaging mode with polarimetric optics (we will call this mode {\it filter polarimetry}, 104 measurements of 81 stars), or in spectropolarimetric mode (51 measurements of 21 stars) or using both techniques at different epochs (for five stars). All observations with filter polarimetry were obtained with the FORS2 instrument \citep{Appetal98} of the ESO VLT, using mainly the R\_SPECIAL filter (six stars were observed also using the v\_high filter). All observations in spectropolarimetric mode were obtained again using FORS2, mostly with the 300V grism or with both the 300V and the 300I grism. Two stars were observed with FORS2 using the 1200B grism, one star with the 600B grism. Among the 97 stars that we have observed in polarimetric mode, 86 had never been observed before with this technique. Some targets that had been observed before were re-observed by us based on various motivations, for instance, to re-observe in spectropolarimetric mode stars that have been found to be magnetic with filter polarimetry.

Data reduction of spectropolarimetry was carried out as explained by \citet{BagLan18}. For imaging polarimetry we used exactly the same technique of {\it aperture polarimetry} adopted for linear polarisation measurements of minor bodies of the solar system described in Sect.~2.2 of \citet{Bagetal16}, except that we measured only one reduced Stokes parameter ($V/I$) instead of two ($Q/I$ and $U/I$). 
%%%%%%%%%%%%%%%%%%%%%%%%%%%%%%%%%%%%%%%%%%%%%%%%%%%%%%%%%%%%%%%%%%%%%%%%%%%%%%%%%%%%%%%%%
\begin{figure}
\begin{center}
\end{center}
\includegraphics[angle=0,width=8cm,trim={0.7cm 6.2cm 2.0cm 4.3cm},clip]{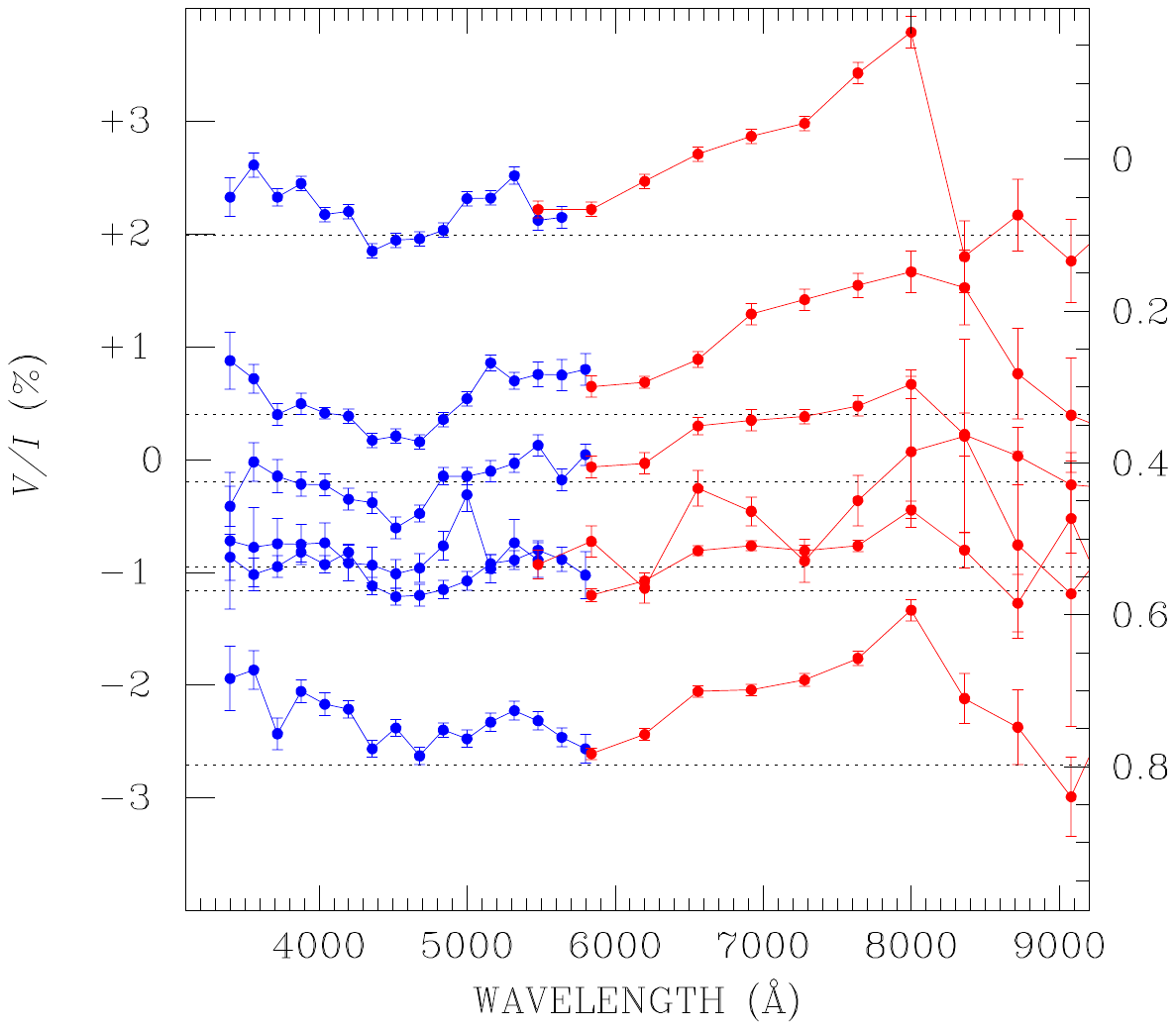}
\caption{\label{Fig_WD0912} Phase-resolved low-resolution circular polarisation spectra of WD\,0912+536, offset proportionally to the rotational phase \citep[shown on the right-hand $y$-axis and calculated according to the zero point given by][and the period by \citeauthor{Faretal24}  \citeyear{Faretal24}]{Angetal72-eph}. For each polarisation spectrum, the zero-level is shown with a dotted line.}
\end{figure}
%%%%%%%%%%%%%%%%%%%%%%%%%%%%%%%%%%%%%%%%%%%%%%%%%%%%%%%%%%%%%%%%%%%%%%%%%%%%%%%%%%%%%%%%%
%%%%%%%%%%%%%%%%%%%%%%%%%%%%%%%%%%%%%%%%%%%%%%%%%%%%%%%%%%%%%%%%%%%%%%%%%%%%%%%%%%%%%%%%%
\begin{figure}
\begin{center}
\end{center}
\includegraphics[angle=0,width=8.8cm,trim={1.1cm 6.0cm 1.1cm 2.8cm},clip]{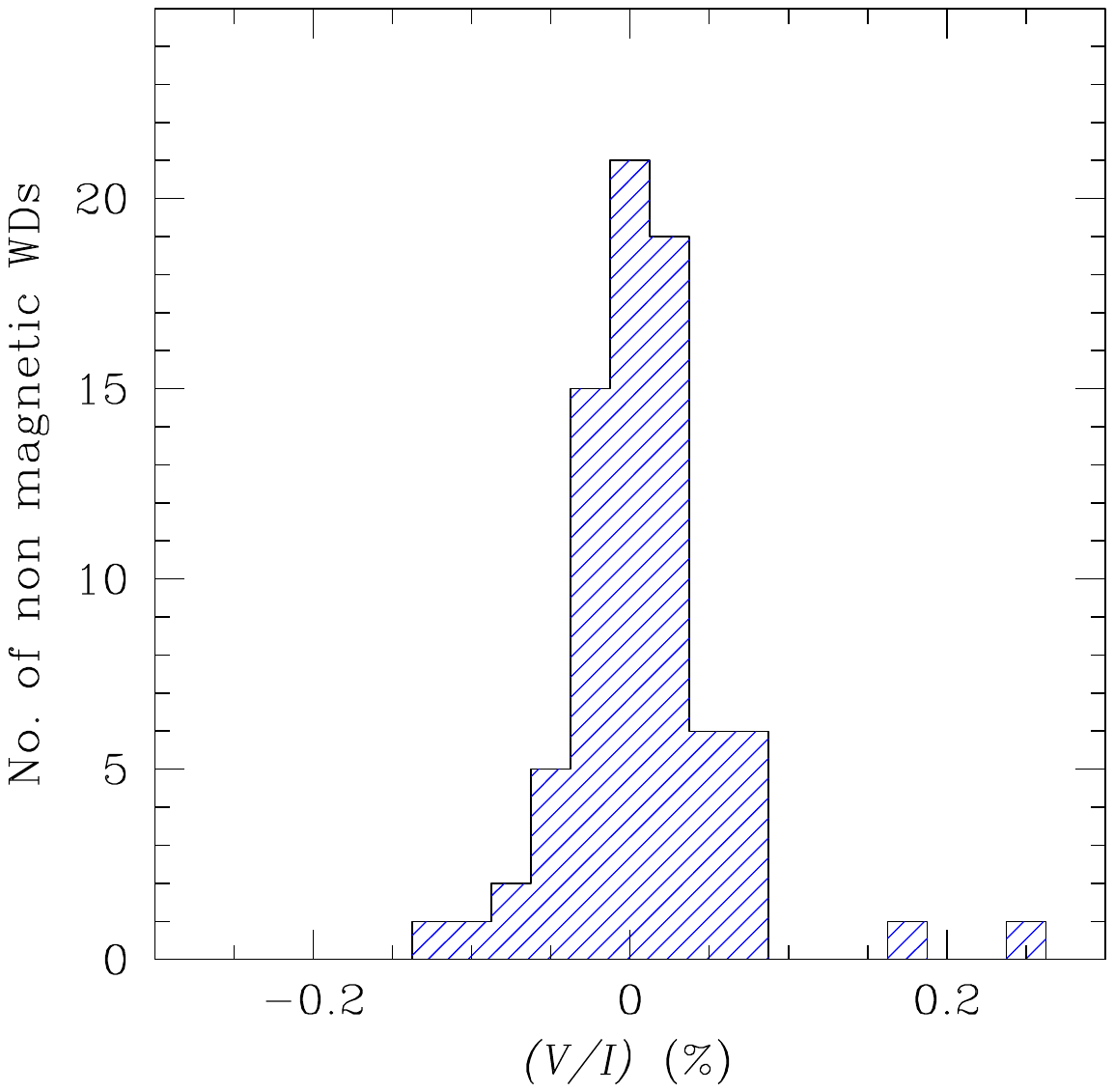}
\caption{\label{Fig_FilterHisto} The distribution of the measured filter polarimetry in the R filter for the stars that have been deemed non-magnetic. }

\end{figure}
%%%%%%%%%%%%%%%%%%%%%%%%%%%%%%%%%%%%%%%%%%%%%%%%%%%%%%%%%%%%%%%%%%%%%%%%%%%%%%%%%%%%%%%%%
%%%%%%%%%%%%%%%%%%%%%%%%%%%%%%%%%%%%%%%%%%%%%%%%%%%%%%%%%%%%%%%%%%%%%%%%%%%%%%%%%%%%%%%%%
\begin{figure}
\begin{center}
\end{center}
\includegraphics[angle=0,width=8.7cm,trim={1.3cm 6.4cm 1.1cm 2.9cm},clip]{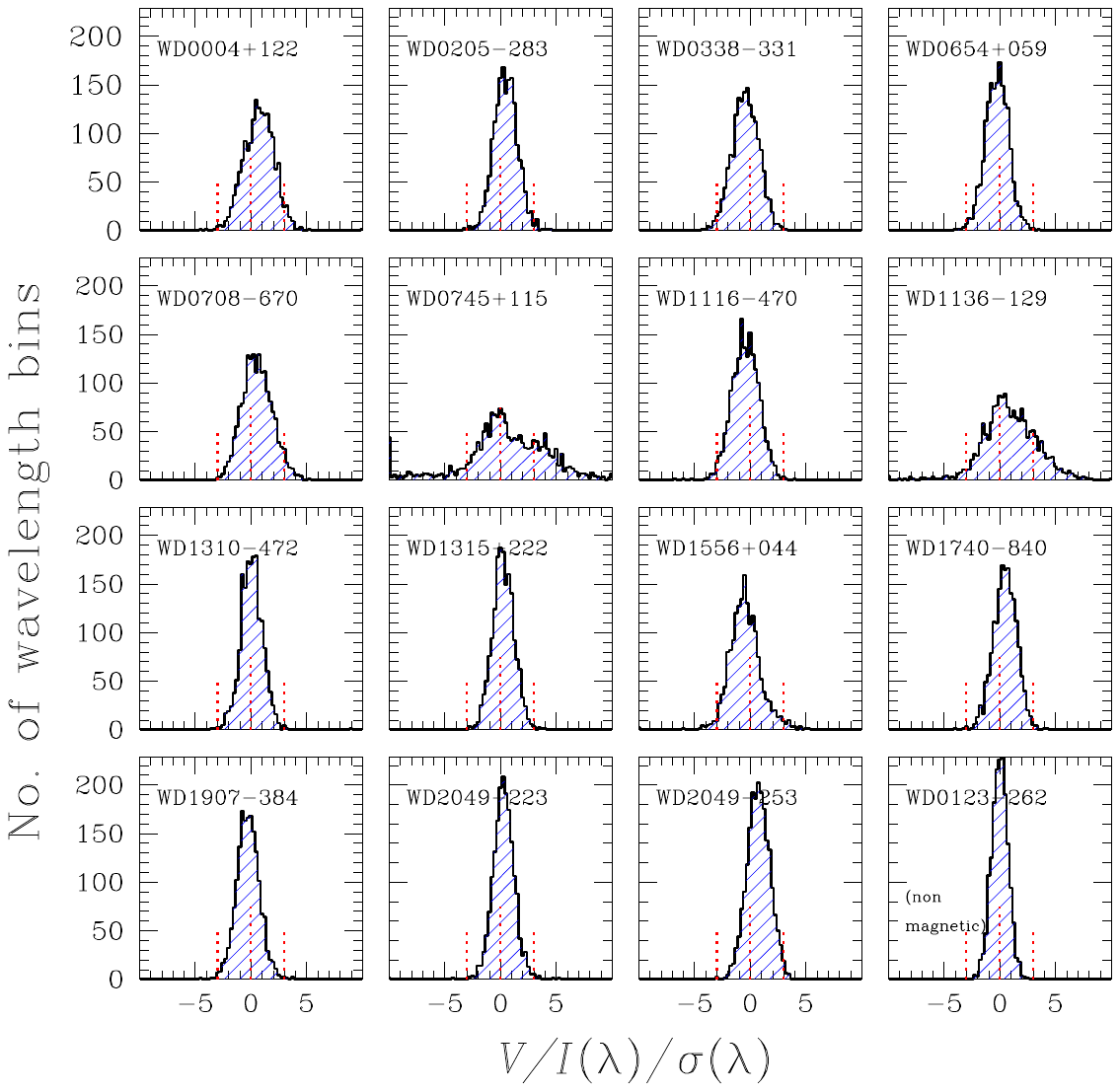}
\caption{\label{Fig_SpectroHisto} The distribution of the $V/I (\lambda)$ values normalised to their uncertainties, for a selection of FORS2 circular polarisation spectra of magnetic white dwarfs, one suspected magnetic (WD\,1310--472), and one non-magnetic white dwarf (WD\,0123--262). The red, vertical dotted lines show the position of the zero and $\pm 3$. }
\end{figure}
%%%%%%%%%%%%%%%%%%%%%%%%%%%%%%%%%%%%%%%%%%%%%%%%%%%%%%%%%%%%%%%%%%%%%%%%%%%%%%%%%%%%%%%%%

The observing log with the polarisation measurements is shown in Table~\ref{Table_Log}. For spectropolarimetric data we reported the value numerically integrated using the filter curves of filters b\_high, v\_high, R\_SPECIAL and I\_BESS, that hereafter will be referred to as b, v, R, and I. For completeness, we also show the log of previous spectropolarimetric measurements obtained with the FORS2 instrument and with ISIS, published by \citet{BagLan18,BagLan20,BagLan21,BagLan24}. 

We also publish for the first time a set of low-resolution spectropolarimetric data of the exceptionally hot and strongly variable DCH WD\,0912$+$536 (Table~\ref{Table_WD0912} and Fig.~\ref{Fig_WD0912}). This star was the second magnetic white dwarf discovered \citep{AngLan71-Second}, and the first magnetic WD found to exhibit variable circular polarisation  \citep{AngLan71-Periodic}. During the 1970s, low resolution circular polarisation spectra were obtained of several strongly polarised magnetic WDs, using the Palomar 5-m telescope with the Oke multi-channel scanner equipped with a polarimetric analyser \citep{Angetal72-GRW,LanAng74,LanAng75,Angetal75}. During this campaign, a series of six circular polarisation spectra, spaced roughly through the 1.33 day rotation and polarisation variation cycle of WD\,0912+536 were obtained. 

In this paper we adopt the definition of the handedness of circular polarisation given by \citet{BagLan19a}. We note that some literature has adopted the opposite definition, so that a comparison between spectra published by different authors must take into account a possible sign ambiguity.

\section{Results}
We re-detected polarisation in 11 stars that were already known from the literature to host magnetic fields. In addition, seven stars were found to be magnetic for the first time in this survey (see Sect.~\ref{Sect_New}), either through filter polarimetry alone, through spectropolarimetry alone, or through a combination of the two techniques.

With filter polarimetry, the only ambiguous cases are WD\,1310$+$027, for which a marginal 3\,$\sigma$ detection led us to classify the star as a suspected magnetic white dwarf, and WD\,1946--735, an ultramassive white dwarf that seems to be rapidly rotating (see Sect.~\ref{Sect_Suspected}). In all other cases, our targets have firm detections, usually confirmed by additional observations, or clear non-detections within uncertainties. Figure~\ref{Fig_FilterHisto} shows a histogram of the signals measured in the stars that we classify as non-magnetic. That histogram is centered about zero (at 0.008\,\%, with 0.05\,\% standard deviation), confirming the absence of any significant instrumental polarisation.

The situation with spectropolarimetry is sometimes more ambiguous. In some cases, it is not immediately clear whether the polarisation spectrum of a star reveals a weak signal or whether the observed signal is consistent either with noise or with a small instrumental polarisation. 

A possible way to assess whether a polarisation spectrum departs from zero is to consider the histogram of the $V/I(\lambda)$ values normalised to their uncertainties. For a non-magnetic star, this histogram is expected to be centred around zero, with an approximately Gaussian shape and $\sigma \simeq 1$ (see Fig.~\ref{Fig_SpectroHisto}). Clear departures from that shape indicate the presence of a polarisation signal, especially if the same behaviour is recovered from more than one dataset. When a small deviation from zero is consistently detected over multiple epochs, it can provide evidence for a genuine magnetic signal rather than a random fluctuation compatible with zero polarisation. An example of a very small signal recovered on multiple occasions is that of WD\,1116--470, shown in Fig.~B.8 of \citet{BagLan24}. The detection of a non-zero polarisation in WD\,1116--470 is also confirmed by multi-filter polarimetry that we obtained with FORS2 (see Table~\ref{Table_Log}). In the cases of WD\,1315$+$222 and WD\,2049--222, our polarisation spectra alone do not yield a firm detection. We therefore base our final assessment on the independent detection of a small polarisation signal obtained through filter polarimetry by \citet{Beretal22,Beretal23,Beretal24}. 

\subsection{Newly discovered magnetic stars}\label{Sect_New}
In the following, we briefly comment on each of the seven newly discovered polarised white dwarfs. 
\subsubsection{WD\,0042--064 = PHL 6585}   
This is a particularly old (9.8\,Gyr) DC white dwarf, with relatively high mass ($\sim 0.93\,M_\odot$). We observed it three times in broadband circular polarisation with the R filter, and all three times we detected a small ($\simeq 0.2$\,\%) signal. The third observing series (obtained on 2024-12-29) consists of four pairs of exposures and the signal measured from different pairs shows a scattering larger than we would expect from photon-noise. This star is therefore a candidate for variability of the order of minutes, that should be assessed with a dedicated time-series. 

\subsubsection{WD\,0205--283 = Gaia DR3 5117287339424622720}
This is another extremely old ($\tau \sim 9.1$\,Gyr) WD with "normal" mass ($\sim 0.69\,M_\odot$) in which we detected a 0.6\,\% polarisation signal. We first discovered its magnetic field from broad-band circular polarisation, then, about two years later, we re-observed it in spectropolarimetric mode, and one month later, once more in filter polarimetry. The measured polarisation in the R filter seems constant with time, hence we assume that the star does not show variability.  

\subsubsection{WD\,0338--331 = [OHD2001] WD\,J0340--330} 
This star was observed in spectropolarimetric mode hence we can confirm the lack of features in its intensity spectrum, at least at the low resolution of the 300V grism. With an estimated cooling age of approximately 10 Gyr, this is one of the oldest known magnetic white dwarfs. It has a relatively large mass ($\sim 0.90\,M_\odot$) and has the strongest apparently continuum-dominated polarisation signal in our sample ($\simeq 2.1$\,\%), which makes this object quite remarkable. 

\subsubsection{WD\,1136--129 = UCAC4 384-059519}    
Previous studies have classified WD\,1136--129 as a featureless He-rich DC \citep{Treetal20,McCetal20,Caretal23}, but our polarised spectrum shows the signature of H lines (middle panels of Fig.~\ref{Fig_WD1556}). The two polarisation spectra that we have obtained differ slightly from each other, and we classify this star as variable. The intensity spectrum shows the presence of spectral lines and the star will be further discussed in Sect.~\ref{Sect_Non_DC}.

\subsubsection{WD\,1434$-$558 = Gaia DR3 5893318763718868736   }%
This featureless He-rich DC has a mass slightly larger than average ($0.725\,M_\odot$), and  we have detected a signal in both the R ($\sim -1.6$\,\%) and the V ($\sim -1.4$\,\%) filters. It has been observed only once, hence we cannot assess its variability.

\subsubsection{WD\,1740--840 = Gaia DR2 5767614206300408576}  
In this normal mass white dwarf ($M \sim 0.66\,M_\odot$), we discovered a magnetic field using filter polarimetry.  We then repeatedly observed it in spectropolarimetric mode with the 300V grism. We did not find evidence of variability.

\subsubsection{WD\,1907--384 = PM J19110-3820}
This is a featureless DC of average mass ($M \sim 0.59\,M_\odot$), and is a rather old ($\tau=8.9$\,Gyr) star. We have repeatedly observed it with the 300V and 300I grisms. We found a small signal of polarisation ($\sim 0.3$\,\%) but no evidence of variability.

\section{Combining new data with literature data}\label{Sect_Combi}
Previous circular polarimetric observations of DC white dwarfs were obtained by 
\citet{AngLan71-Second,AngLan71-Periodic} and \citet{Angetal72-eph} (spectropolarimetry of WD\,0912+536); 
\citet{LieSto80} and \citet{Angetal81} (broadband polarimetry of a few DC WDs);
\citet{Putney95,Putney97} (the first systematic spectropolarimetric survey of DC WDs);
\citet{Schetal01} (spectropolarimetry of WD\,0000--345); 
\citet{BagLan18,BagLan20,BagLan21,BagLan24} (spectropolarimetry of a few DC WDs); and
\citet{Beretal22,Beretal23,Beretal24} (multifilter polarimetry of about 80 DC WDs).

Before the present survey, 17 objects in our DC-selected compilation were known to be magnetic. Five magnetic DC white dwarfs were found in the local 20\,pc volume: 
WD\,0004+122, 
WD\,0708--670, 
WD\,2049--253 \citep{BagLan20}, all three discovered via spectropolarimetry;
WD\,0912+536 \citep[][via multifilter polarimetry]{AngLan71-Second}; and
WD\,1116--470 \citep{BagLan21,BagLan24}, again via spectropolarimetry. 
Outside the local 20\,pc volume, 
WD\,0756+437 was discovered by \citet{Putney95} via spectropolarimetry,
WD\,1556+044 and 
WD\,2049--222 were discovered by \citet{Beretal22}, although WD\,1556+044 was later revealed to be a DA (see Sect.~\ref{Sect_Non_DC}); 
five more stars, 
WD\,1315+222, 
WD\,1346+121, 
WD\,1601--073, 
WD\,1853+775, and 
WD\,2211+372
were discovered by \citet{Beretal23};  
and four more,
WD\,0004+337, 
WD\,0419+576,
WD\,0654+059,
WD\,0745+115
by \citet{Beretal24}. We note that WD\,0019+423, WD\,0052+595, and WD\,0021+683 were reported to be magnetic by \citet{Beretal24}, but a reanalysis of the observations by Stroet et al. (2026, in preparation) showed that none of these stars exhibit a significant polarisation signal. All discoveries by \citet{Beretal22,Beretal23,Beretal24} were obtained via multi-filter polarimetry. For some of these stars, we also obtained FORS polarisation spectra 
(WD\,0654+059, 
WD\,0745+115, 
WD\,1315+222, 
WD\,1346+121, 
WD\,1601--073,
WD\,2049--222). In addition, the survey by \citet{KawVen12} mentions FORS1 spectropolarimetry of five DC white dwarfs, all between 20 and 40\,pc. Our examination of the data suggests that one of them, WD\,0434--611 may show a weak signal of polarisation, and will be treated as suspected magnetic.
Spectropolarimetric observations are available for 17 of the confirmed magnetic objects in our compilation; some of these show identifiable contributions from spectral lines, as discussed in Sect.~\ref{Sect_Non_DC}.

To the best of our knowledge, the only linear polarisation measurement of a magnetic DC star is that from \citet{Angetal81} for WD\,0912$+$536 (=EGGR\,250=G195-19), resulting in a non-detection.

Table~\ref{Table_Stars} collects the information about all DC white dwarfs observed in polarimetric mode that are known to us. Stellar parameters for the stars within 40\,pc are from \citet{OBretal24}, the remaining ones are obtained using the effective temperature and mass from \citet{Genetal21} and cooling tables of \citet{Bedetal20}. Table~\ref{Table_Stars} gives the maximum of the absolute value of the polarisation measured in that star, the filters or grisms in which polarimetric data were acquired, and a reference to the published measurements. Table~\ref{Table_MWD} reports the same data only for the magnetic, or suspected magnetic white dwarfs.
%%%%%%%%%%%%%%%%%%%%%%%%%%%%%%%%%%%%%%%%%%%%%%%%%%
\begin{table}
\caption{\label{Table_MWD} List of 24 confirmed and four suspected magnetic DC-selected white dwarfs with their essential parameters.
}
\begin{center}
\tabcolsep=0.05cm
\begin{tabular}{crrrcll}
\hline \hline
      \multicolumn{1}{c}{STAR}          & 
      \multicolumn{1}{c}{$T_{\rm eff}$} &
      \multicolumn{1}{c}{$M$}           &
      \multicolumn{1}{c}{$\tau$}        &
      \multicolumn{1}{c}{$\vert V/I\vert_{\rm max}$} &
      &
      \multicolumn{1}{l}{REFERENCES} \\
                                       &
     \multicolumn{1}{c}{(K)}           &
     \multicolumn{1}{c}{($M_\odot$)}   &
     \multicolumn{1}{c}{ (Gyr)} &
     \multicolumn{1}{c}{(\%)}  & 
     &
     \\
     \hline
WD\,0004$+$122    &5285   &0.81  &  7.83    &0.9 &\splc& BL20, \tw\ (Fig.~\ref{Fig_Spectra})\\ 
WD\,0004$+$337    &5666   &0.71  &  5.26    &0.6 &   & Be+24 \\ 
WD\,0042$-$064    &4938   &0.93  &  9.79    &0.3 &v: & \tw \\ 
WD\,0205$-$283    &4701   &0.69  &  9.12    &1.2 &nv & \tw\ (Fig.~\ref{Fig_Spectra})\\ 
WD\,0338$-$331    &4844   &0.90  & 10.02    &2.1 &v: & \tw\ (Fig.~\ref{Fig_Spectra})\\ 
WD\,0419$+$576    &7140   &0.69  &  2.08    &0.2 &   & Be+24 \\ 
(WD\,0434$-$611)  &4861   &0.63  &  7.68    &0.1:&   & KV12  \\ 
WD\,0654$+$059    &6010   &0.69  &  3.97    &0.5 &nv \splc & Be+24, \tw\ (Fig.~\ref{Fig_WD1556})\\ 
WD\,0708$-$670    &5007   &0.59  &  6.18    &0.6 &nv & BL20, \tw\ (Fig.~\ref{Fig_Spectra})\\ 
WD\,0745$+$115    &9831   &0.93  &  1.71    &1.9 &v \spl & Be+24 \\  
WD\,0756$+$437    &7215   &1.04  &  4.44    &8.5 &v \spl &Ptn95  \\  
WD\,0912$+$536    &7155   &0.71  &  2.26    &1.0 &v  & AL71, \tw\ (Fig.~\ref{Fig_WD0912})\\  
WD\,1116$-$470    &5903   &0.62  &  3.36    &0.2 &nv & BL21, \tw\ (Fig.~\ref{Fig_Spectra})\\ 
WD\,1136$-$129    &5930   &0.65  &  3.80    &1.5 & v\ \spl & \tw\ (Fig.~\ref{Fig_WD1556})\\ 
(WD\,1310$-$472)  &4097   &0.61  &  9.75    &0.1:&   & BL21 (Fig.\ref{Fig_WD1310}) \\
(WD\,1310$+$027)  &4384   &0.64  &  9.45    &0.1:&   & Be+24, \tw  \\
WD\,1315$+$222    &6334   &0.73  &  3.54    &0.3 &nv & Be+23, \tw\ (Fig.~\ref{Fig_Spectra})\\ 
WD\,1346$+$121    &4607   &0.74  &  8.10    &1.2 &nv & Be+23, \tw\ \\ 
WD\,1434$-$558    &8210   &0.73  &  1.53    &1.6 &   & \tw  \\ 
WD\,1556$+$044    &6730   &0.85  &  3.81    &0.6 &nv \spl & Be+22, \tw\ (Fig.~\ref{Fig_WD1556})\\ 
WD\,1601$-$073    &4995   &1.03  &  9.37    &1.6 &   & Be+23, \tw\\ 
WD\,1740$-$840    &5123   &0.67  &  6.52    &0.8 &nv & \tw\ (Fig.~\ref{Fig_Spectra})\\ 
WD\,1853$+$775    &5211   &0.57  &  5.15    &0.7 &   & Be+23\\
WD\,1907$-$384    &4328   &0.59  &  8.89    &0.4 &nv & \tw\ (Fig.~\ref{Fig_Spectra})\\ 
(WD\,1946$-$735)  &4934:  &1.28: &  5.51:   &1.5:&v: & \tw  \\
WD\,2049$-$253    &5048   &0.58  &  5.70    &0.6 &nv & BL20, \tw\ (Fig.~\ref{Fig_Spectra})\\ 
WD\,2049$-$222    &7811   &0.69  &  1.60    &0.1 &   & Be+22, \tw\ (Fig.~\ref{Fig_Spectra})\\ 
WD\,2211$+$372    &6424   &0.88  &  4.37    &1.3 &   & Be+23 \\
\hline
\end{tabular}
\end{center}
   \begingroup
    \let\clearpage\relax 
    {\bf Notes:} Stars whose names are in parentheses are suspected magnetic --  detection
should be confirmed (or ruled out) by further observations. Label \spl\ refers to stars whose spectra
show presence of spectral lines. Label "nv" means that the star does not show evidence of variability; 
"v:" means suspected variable; 
"v" means that the polarisation of the star is definitely variable with time.\
The last column lists the references in which polarisation measurements are reported for the first time, as follows:
\tw: this work;
AL71: \citet{AngLan71-Second}, \citet{AngLan71-Periodic};
Be+22: \citet{Beretal22};
Be+23: \citet{Beretal23};
Be+24: \citet{Beretal24};
BL18: \citet{BagLan18};
BL20: \citet{BagLan20};
BL21: \citet{BagLan21};
BL24: \citet{BagLan24};
KV12: \citet{KawVen12};
Ptn95: \citet{Putney95}. The same columns report our new observations (\tw), regardless of whether they represent a first detection.
References to additional measurements are given in Table~\ref{Table_Stars}.
     \endgroup
\end{table}

%%%%%%%%%%%%%%%%%%%%%%%%%%%%%%%%%%%%%%%%%%%%%%%%%%
\subsection{Suspected magnetic DC stars}\label{Sect_Suspected}
%%%%%%%%%%%%%%%%%%%%%%%%%%%%%%%%%%%%%%%%%%%%%%%%%%%%%%%%%%%%%%%%%%%%%%%%%%%%%%%%%%%%%%%%%
\begin{figure}
\begin{center}
\includegraphics*[angle=0,width=8.6cm,trim={0.5cm 6.5cm 1.0cm 2.8cm},clip]{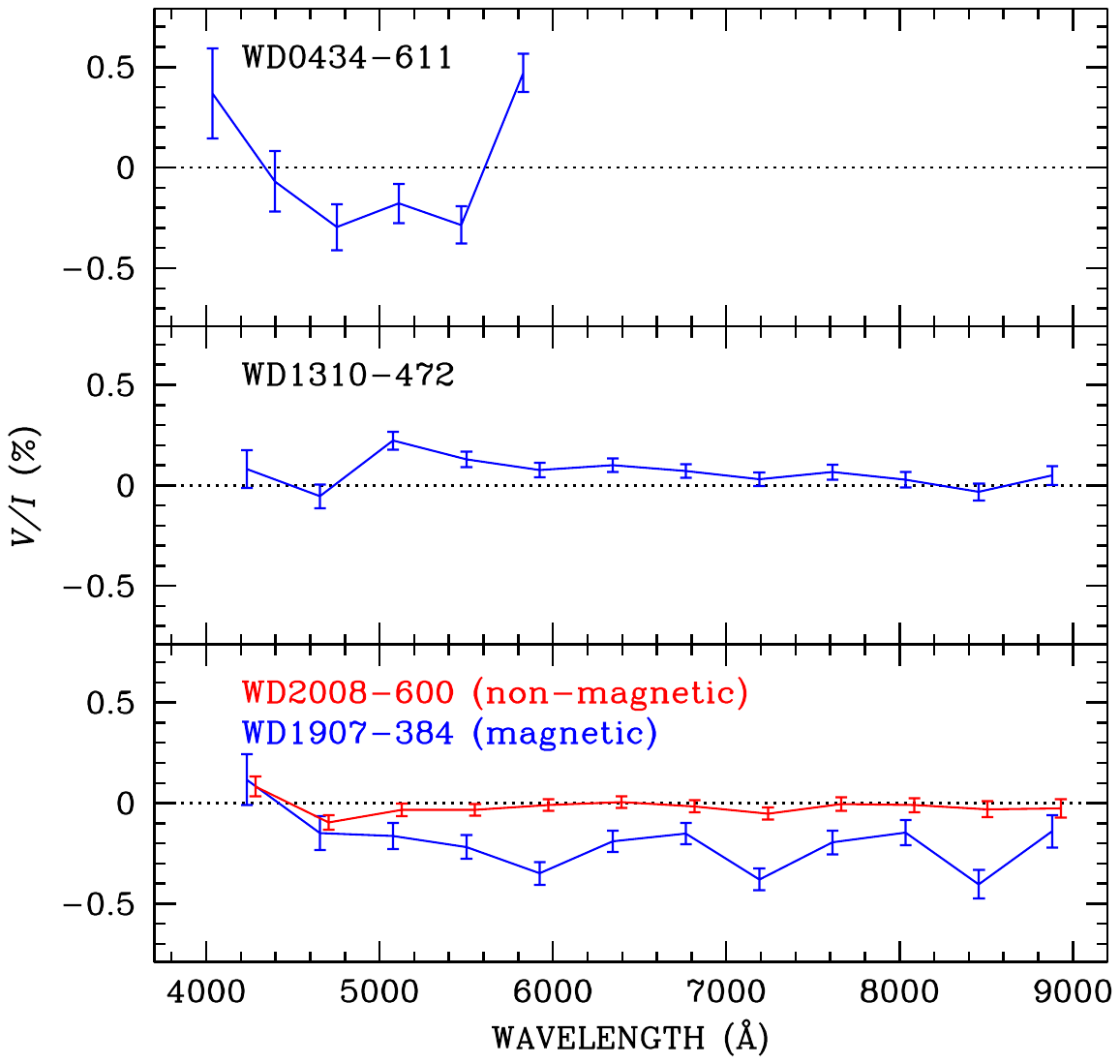}
\end{center}
\caption{\label{Fig_WD1310} Circular polarisation spectra of two suspected magnetic stars, WD\,0434--611 and WD\,1310--472. For comparison, spectra of the weakly but firmly polarised WD\,1907--384 and the non-polarised WD\,2008--600 are also shown.}
\end{figure}
%%%%%%%%%%%%%%%%%%%%%%%%%%%%%%%%%%%%%%%%%%%%%%%%%%%%%%%%%%%%%%%%%%%%%%%%%%%%%%%%%%%%%%%%%
Four DC white dwarfs remain as suspected magnetic stars.
%WD\,0052+595, %Be+24
%WD\,1310$-$427, %300V, BL21
%WD\,1310$+$027, and %Be+ and this work
%WD\,1946--735. %this work
One of them, WD\,1310--472, exhibits a small but systematic departure from zero polarisation in its spectrum, as previously discussed by \citet{BagLan21}. WD\,0434--611 may show a weak polarisation signals. Since the detection remains marginal and both stars have been observed only once (in spectropolarimetric mode), we classify them as suspected magnetic. Their polarisation spectra are shown in Fig.~\ref{Fig_WD1310}, together, for comparison, with the spectra of a firmly detected magnetic star (WD\,1907--384) and a non-polarised white dwarf (WD\,2008--600).

The two remaining suspected magnetic stars are such because they show marginal detection with filter polarimetry. These are WD\,1310$+$027, and 
WD\,1946--735. 

For WD\,1310$+$027 \citet{Beretal24} did not detect any significant continuum polarisation, reporting an uncertainty of approximately 0.08\,\%. Our new FORS2 observations, instead, yield a marginal detection of $0.11 \pm 0.04$\,\%, corresponding to a nearly $3\,\sigma$ detection. A polarisation signal at the $\sim 0.1$\,\% level would not have been detectable in the measurements of \citet{Beretal24}. Given the limited significance of our result, the FORS2 observations do not provide conclusive evidence that the star is intrinsically polarised.

According to the catalogue of \citet{Genetal21}, WD\,1946--735 is an ultra-massive white dwarf with $M \simeq 1.28\,M_\odot$. However, \citet{OBretal23} noted evidence of weak collision-induced absorption (CIA) in its optical spectrum. Because only broad-band Gaia photometry is available for this star, the degeneracy between $T_{\rm eff}$ and the atmospheric hydrogen-to-helium abundance ratio, $\log({\rm H/He})$, cannot be resolved, preventing a reliable determination of its physical parameters. We obtained two sequences of broadband circular polarimetry on the same night. The first sequence consisted of two beam-swapping pairs, with individual exposures of 664\,s, and yielded no significant detection of circular polarisation. The second sequence comprised four beam-swapping pairs with shorter individual exposures of 340\,s. Combining all exposures from this sequence gives a marginal $3\sigma$ detection of $V/I = 0.24 \pm 0.08$\,\%. A closer inspection of this second dataset reveals that the circular polarisation derived from the individual beam-swapping pairs varies beyond what is expected from photon noise. In particular, the measurements span values from approximately $+1.5$\,\% to $-0.3$\,\%, whereas the typical uncertainty of an individual measurement is only $\sim 0.2$\,\%. Such scatter is difficult to explain as statistical noise alone and suggests that the polarisation may be intrinsically variable on timescales comparable to the duration of the observing sequence. One possible interpretation is that WD\,1946--735 is a rapidly rotating magnetic white dwarf. If the circular polarisation changes significantly during the integration, the longer 664\,s exposures would average over the rotational modulation and substantially reduce the measured signal, naturally explaining the non-detection in the first sequence. At present, however, the available data are insufficient to confirm our interpretation, and additional time-resolved data are required to establish if the star is both magnetic and rapidly rotating.

\subsection{Stars in which spectral lines affect the polarisation spectrum}\label{Sect_Non_DC}
Our target list was prepared by searching the literature for DC white dwarfs. However, 
it is well established that white dwarfs initially classified as featureless sometimes reveal spectral features when observed at higher signal-to-noise ratio or higher spectral resolution, leading to a revision of their spectral classifications. This, for instance, occurred frequently in the survey of DC white dwarfs carried out by \citet{Putney97}. Revisiting the spectral classification may not necessarily appear as an interesting exercise in this context, except that it may be useful to understand whether the shape of the continuum polarisation is affected by spectral lines, or if it is entirely dominated by magnetic dichroism in the continuum opacity. For that reason, it is worthwhile to examine more closely additional cases in which $V/I$ shows the line profiles.

In six stars, the presence of spectral lines was revealed in the $V/I$ spectrum. In particular, the most strongly polarised star in our original sample, WD\,0756+437 was repeatedly classified as DC in the literature. However, the star shows polarisation features due to the presence of various spectral lines, as discussed in the discovery paper by \citet{Putney95}.

\begin{figure*}
\begin{center}
\includegraphics*[angle=0,width=5.8cm,trim={1.0cm 6.0cm 1.0cm 3.0cm},clip]{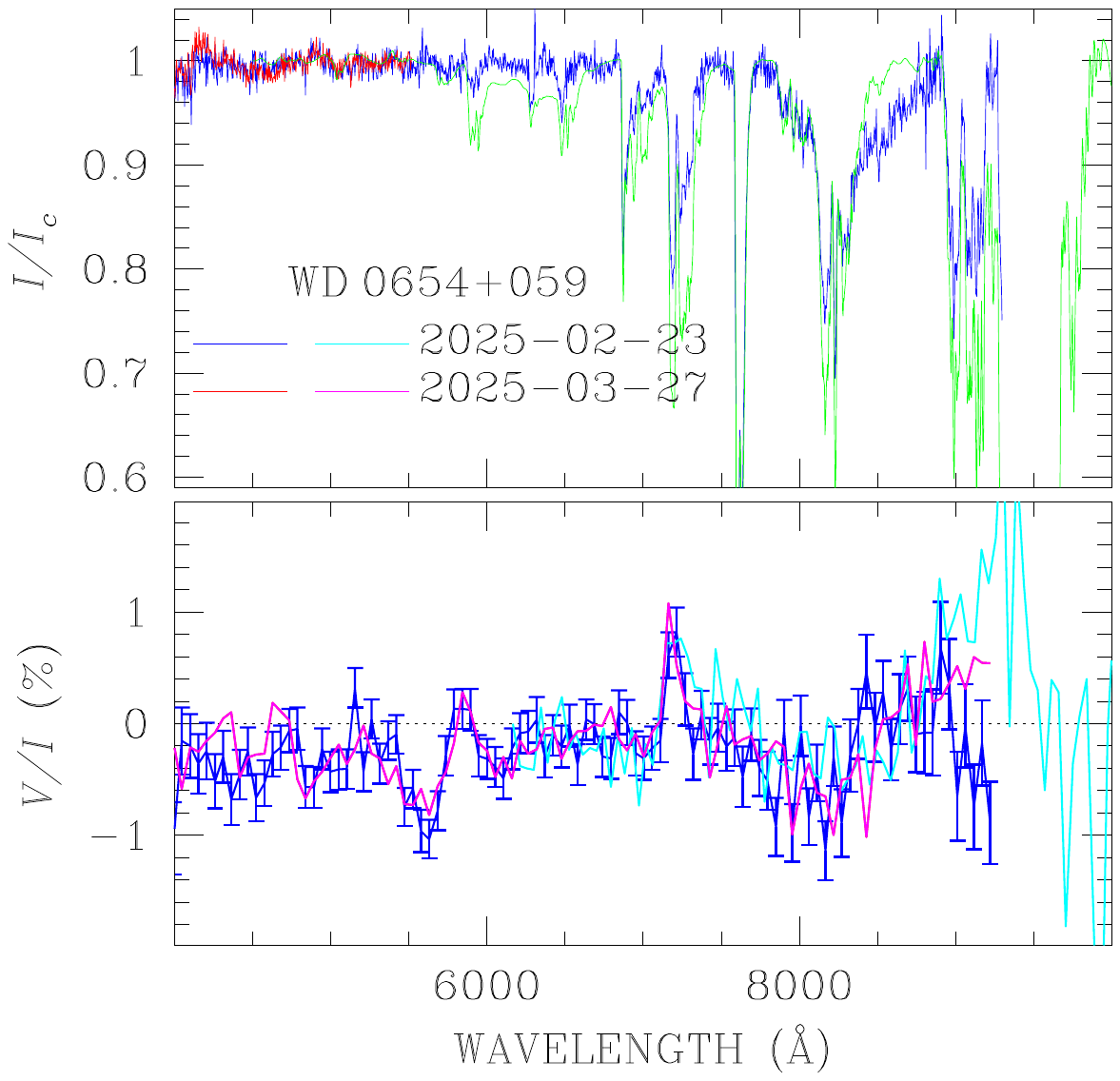}
\includegraphics*[angle=0,width=5.8cm,trim={1.0cm 6.0cm 1.0cm 3.0cm},clip]{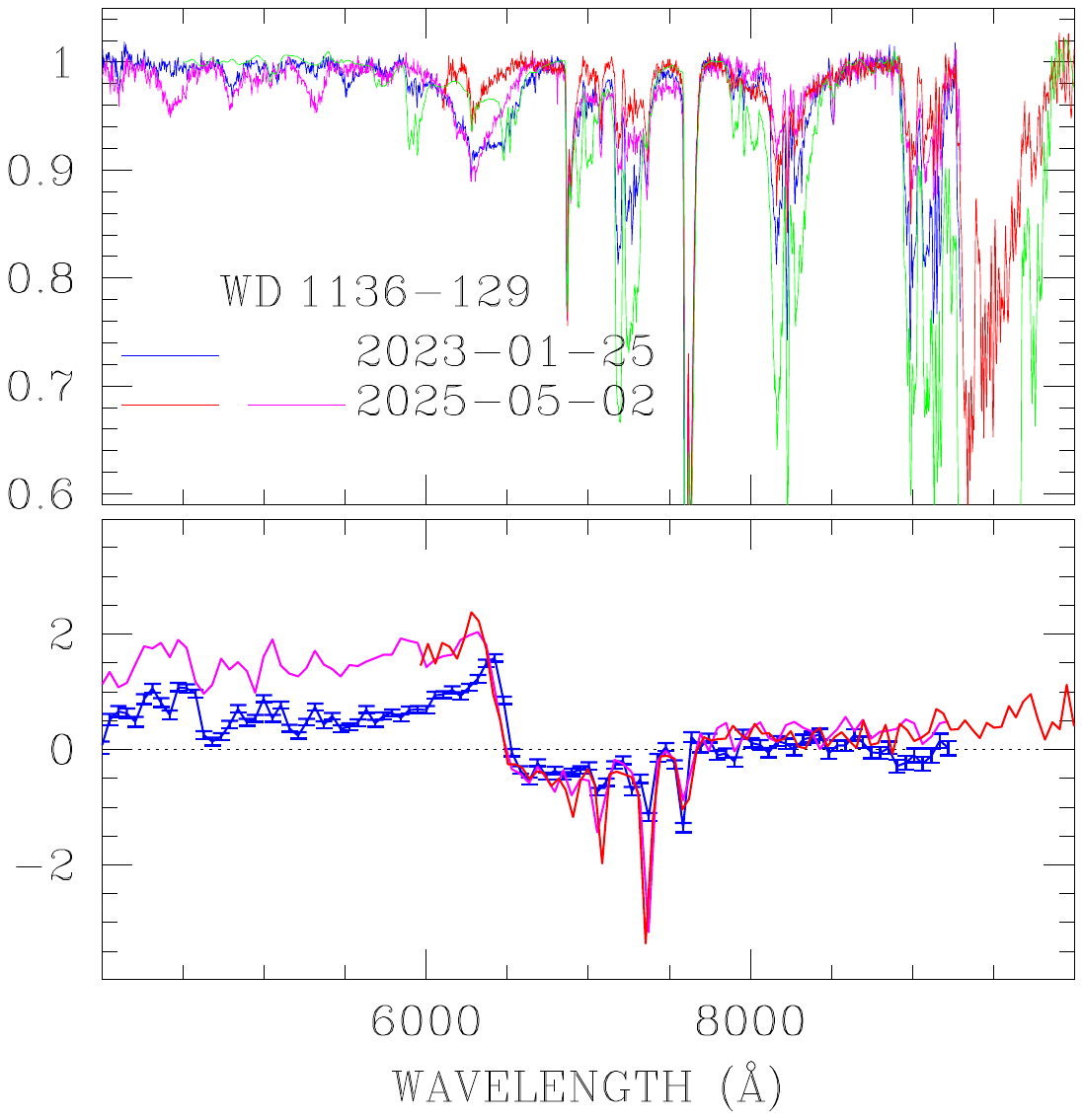}
\includegraphics*[angle=0,width=5.8cm,trim={1.0cm 6.0cm 1.0cm 3.0cm},clip]{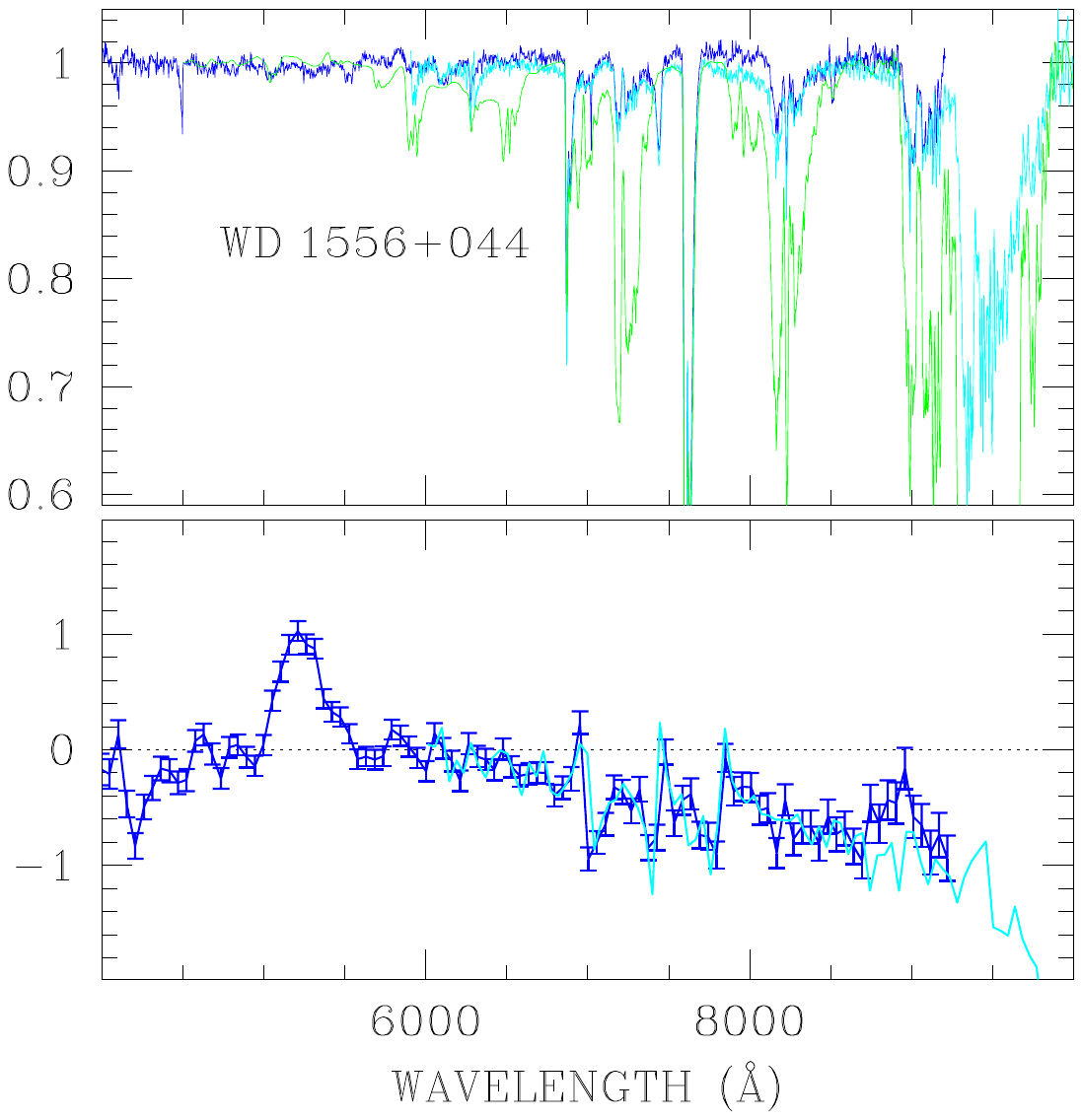}
\end{center}
\caption{\label{Fig_WD1556} 
Flux normalised to the pseudo-continuum $I_{\rm c}$ (from which telluric lines were not removed) and reduced circular polarisation spectrum $V/I$ of the magnetic white dwarfs WD\,0654+059 (observed with both 300V and 300I grisms), WD\,1136--129 (observed on 2023-01-25 with the 300V grism and on 2025-05-02 with both the 300V and 300I grisms) and WD\,1556+044 (obtained with the 300V and 300I grisms). In the top panels, the green curves show the telluric transmission predicted by the ESO FORS2 Exposure Time Calculator.}
\end{figure*}

WD\,0654+059 was classified as He-rich DC white dwarf, and broadband circular polarisation was detected by \citet{Beretal24}. From our spectra we confirm the presence of a small signal of continuum polarisation, but the polarised $\sigma$ components of H$\alpha$ can be detected in two independent observations (see left panels of Fig.~\ref{Fig_WD1556}). Because the line effects are quite tiny, this star will be retained in our sample of DC white dwarfs.

WD\,0745+115 was classified as featureless He-rich star \citep[e.g.][]{OBretal24}, but Stroet et al.\ (in prep.) have shown that it is actually H-rich, and that many of its polarised features are actually caused by H absorption lines. Spectra will be reported by Stroet et al.\ (in prep.).
WD\,1136--129 was classified as DC by \citet{Treetal20}, but our $V/I$ profiles show clearly the presence of spectral lines (see mid panels of Fig.~\ref{Fig_WD1556}).  Similarly, WD\,1556+044 was classified DC in some of the previous literature \citep[e.g.][]{McCetal20,Caretal23}. Obviously, previous filter polarimetry \citep{Beretal22} could not be used to re-assess spectral classification, but our new spectropolarimetry clearly reveals the presence of H lines (see right panels of Fig.~\ref{Fig_WD1556}). In fact, the $\pi$ component of H$\alpha$ can be seen in the intensity spectrum and indeed the star had been classified as magnetic DA by \citet{Kiletal25}. These three stars will no longer be considered as genuine DC white dwarfs.

A sixth potentially similar case is that of the magnetic WD\,0004+122, in which weak H$\alpha$ signatures were identified in the $V/I$ profiles by \citet{BagLan20}. As for WD\,0654+059, since the line effect seems quite small, WD\,0004+122 will be retained in our sample of DC white dwarfs.

%%%%%%%%%%%%%%%%%%%%%%%%%%%%%%%%%%%%%%%%%%%%%%%%%%%%%%%%%%%%%%%%%%%%%%%%%%%%%%%%%%%%%%%%%
%%%%%%%%%%%%%%%%%%%%%%%%%%%%%%%%%%%%%%%%%%%%%%%%%%%%%%%%%%%%%%%%%%%%%%%%%%%%%%%%%%%%%%%%%
WD\,0149-723 was repeatedly classified as DC, but \citet{Kiletal26} claimed that it is a DAZ. Since our spectra do not confirm the presence of either H$\alpha$ or of metal lines we continue to consider it a DC white dwarf.

Our target list included eight white dwarfs outside the local 40\,pc volume that are listed in the catalogue of white dwarfs by \citet{Genetal21}, but whose spectral type could not be established because no spectral classification was available in the literature.  Two of these stars
WD\,0205--283 and WD\,1818--510 were in fact observed by us in spectropolarimetric mode and no obvious line features were identified in their spectra; they are therefore DC WDs. The remaining six stars (WD\,0649--698, WD\,0706-175, WD\,1010--104, WD\,1048--737, WD\,1308--346, WD\,1825--607) have all $T_{\rm eff} \le 5000$\,K, but were observed with filter polarimetry, hence we are not able to reach any conclusion about the presence of spectral lines. In Table~\ref{Table_Stars}, they are marked with a label "(DC?)" in the last column.

In summary, the polarised spectra of 
WD\,0745+115, 
WD\,0756+437, 
WD\,1136--129, 
WD\,1556+044 
should not be considered as pure examples of how a magnetic field affects the propagation matrix of the continuum radiation, as spectral lines contribute to the shape of their $V/I$ profiles. All these stars are labelled with \spl\ in Table~\ref{Table_Stars} and in Table~\ref{Table_MWD}. WD\,0004+122 and WD\,0654+059, which show small features of polarised lines in their spectra, will be retained in the list, and flagged with \splc\ in Table~\ref{Table_Stars} and in Table~\ref{Table_MWD}.

\subsection{The circular polarised spectra of magnetic DC white dwarfs}\label{Sect_Spectra}
%%%%%%%%%%%%%%%%%%%%%%%%%%%%%%%%%%%%%%%%%%%%%%%%%%%%%%%%%%%%%%%%%%%%%%%%%%%%%%%%%%%%%%%%%
\begin{figure}
\begin{center}
\includegraphics[angle=0,width=8.8cm,trim={1.0cm 2.0cm 1.0cm 1.0cm},clip]{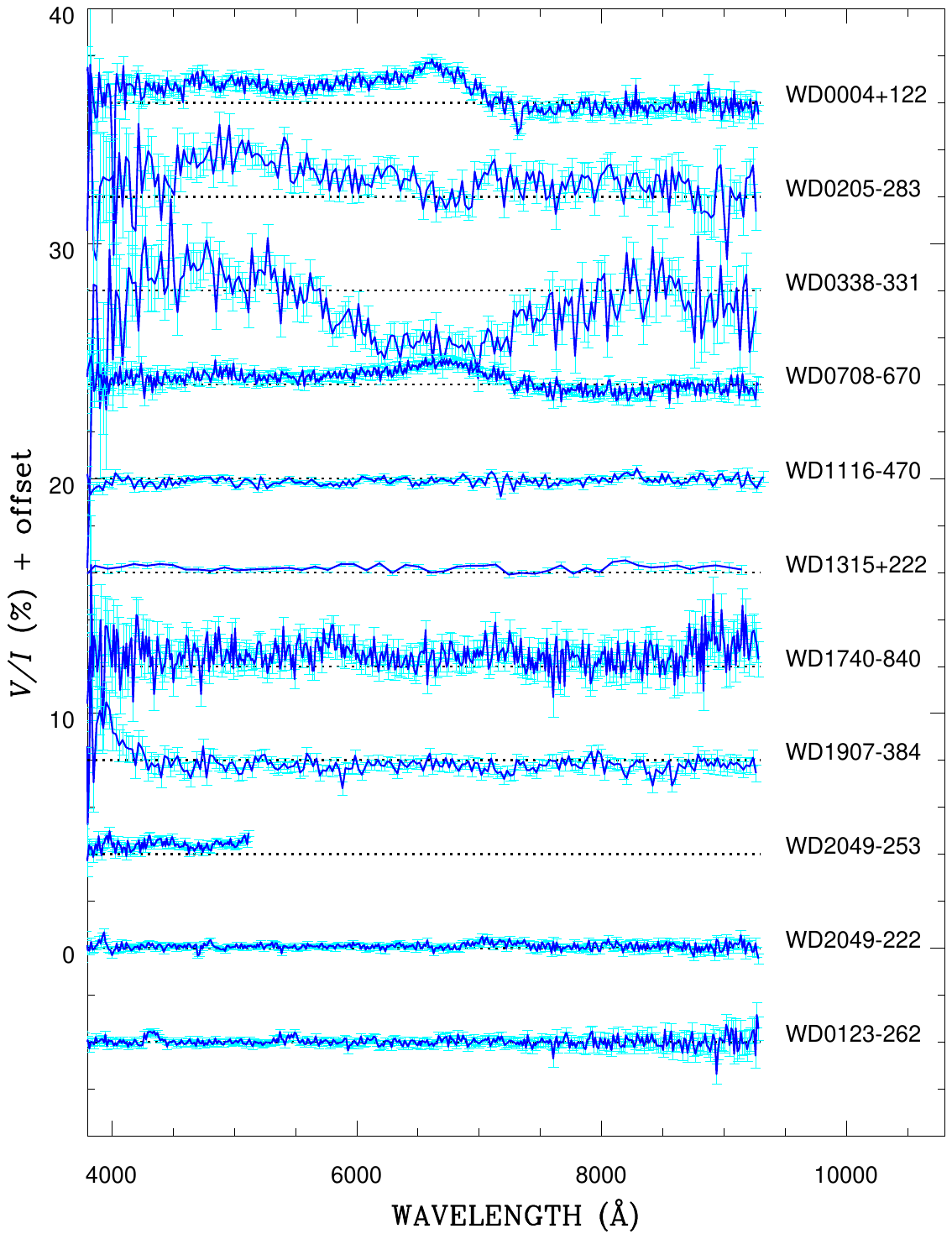}
\end{center}
\caption{\label{Fig_Spectra} 
Circular-polarisation spectra of 10 magnetic DC white dwarfs and one non-magnetic white dwarf (WD\,0123--262) observed with FORS2. The spectra are vertically shifted for display; dotted horizontal lines mark the zero-polarisation level for each star. All spectra were obtained with the 300V grism, except for WD\,2049--253, which was observed with the 1200B grism. Spectra are plotted with different bin sizes, according to the signal-to-noise. In several cases, the polarisation signals are weak, but their detection is confirmed by repeated observations.}
\end{figure}
%%%%%%%%%%%%%%%%%%%%%%%%%%%%%%%%%%%%%%%%%%%%%%%%%%%%%%%%%%%%%%%%%%%%%%%%%%%%%%%%%%%%%%%%%
Figure~\ref{Fig_Spectra} shows a selection of the circular polarisation spectra of the magnetic DC white dwarfs obtained with FORS2. The spectra are vertically shifted for display, and the dotted horizontal lines mark the corresponding zero-polarisation levels. For comparison, the bottom line of the figure shows data for the non-magnetic DC star WD\,0123--262, which is consistent with zero polarisation over the full observed wavelength range. Figure~\ref{Fig_Spectra} does not include WD\,0745+115, WD\,0756+437, WD\,1136--129 and WD\,1556+044 because they are not representative of continuum polarisation, but includes WD\,0004+122 because we think that the contribution to $V/I$ from spectral lines is very small. Spectra of WD\,0912+536 were already shown in Fig.~\ref{Fig_WD0912}. We note that the spectra of WD\,0004+122 and WD\,0708--670 are quite similar. The remaining magnetic DCs generally display a very weak signal, except for WD\,0338--331, the DC with the strongest signal (with the exception of the line-contaminated star WD\,0756+437), in which circular polarisation varies strongly with wavelength and reaches amplitudes of order one per cent or more. Other stars, such as WD\,0654+059, WD\,0708--670, WD\,1116--470, WD\,1315$+$222, and WD\,1907$-$384, show a barely detectable signal.
The diversity of spectral shapes presumably reflects differences in magnetic-field strength, field geometry, and viewing angle, and possibly the effects of variations of composition on continuum opacity. 

\section{Analysis}\label{Sect_Analysis}

%%%%%%%%%%%%%%%%%%%%%%%%%%%%%%%%%%%%%%%%%%%%%%%%%%%%%%%%%%%%%%%%%%%%%%%%%%%%%%%%%%%%%%%%%
\begin{figure}
\begin{center}
\includegraphics[angle=0,width=8.8cm,trim={1.2cm 6.1cm 0.5cm 2.8cm},clip]{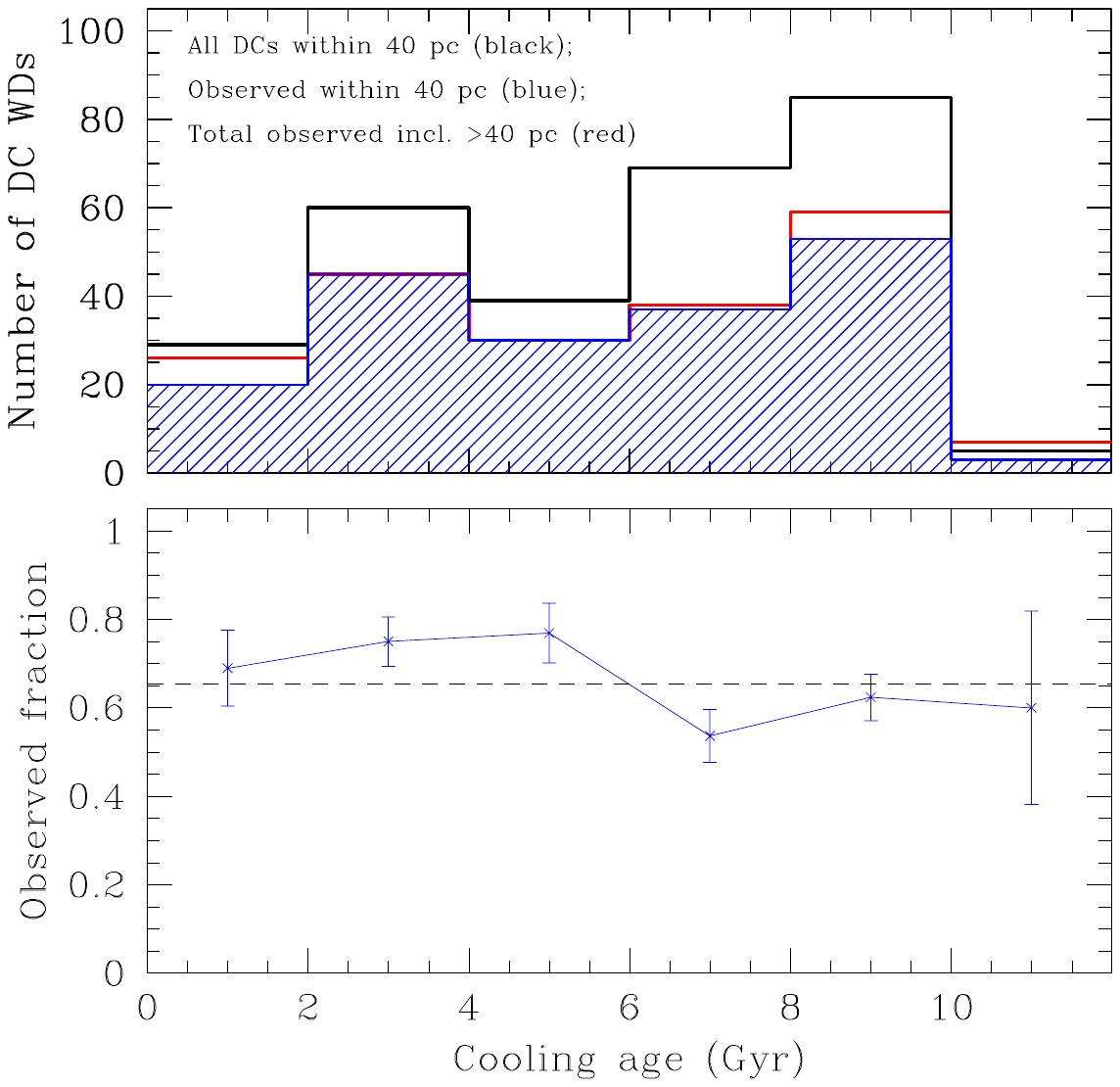}
\end{center}
\caption{\label{Fig_Compli} 
Cooling-age distribution and observational coverage of DC white dwarfs. {\it Upper panel:} Distribution of cooling ages for the 287 DC white dwarfs within 40 pc (black histogram), compared with the 188 stars for which polarimetric observations are available (blue histogram). The red histogram shows the total observed sample, including 17 additional white dwarfs beyond 40\,pc, some of which were deliberately targeted to improve the representation of older objects. {\it Lower panel:} Fraction of DC white dwarfs within 40 pc with available polarimetric observations as a function of cooling age. Error bars represent $1\,\sigma$ binomial uncertainties, while the horizontal dashed line indicates the overall observed fraction  ($\sim 65$\,\%). Both panels use age bins of 2\,Gyr. The additional targets beyond 40\,pc are excluded from the calculation of observational completeness, which refers exclusively to the volume-limited sample.}
\end{figure}
%%%%%%%%%%%%%%%%%%%%%%%%%%%%%%%%%%%%%%%%%%%%%%%%%%%%%%%%%%%%%%%%%%%%%%%%%%%%%%%%%%%%%%%%%
%%%%%%%%%%%%%%%%%%%%%%%%%%%%%%%%%%%%%%%%%%%%%%%%%%%%%%%%%%%%%%%%%%%%%%%%%%%%%%%%%%%%%%%%%%
\begin{figure*}
\begin{center}
\includegraphics[angle=0,width=18cm,trim={1.5cm 11.0cm 0.5cm 2.8cm},clip]{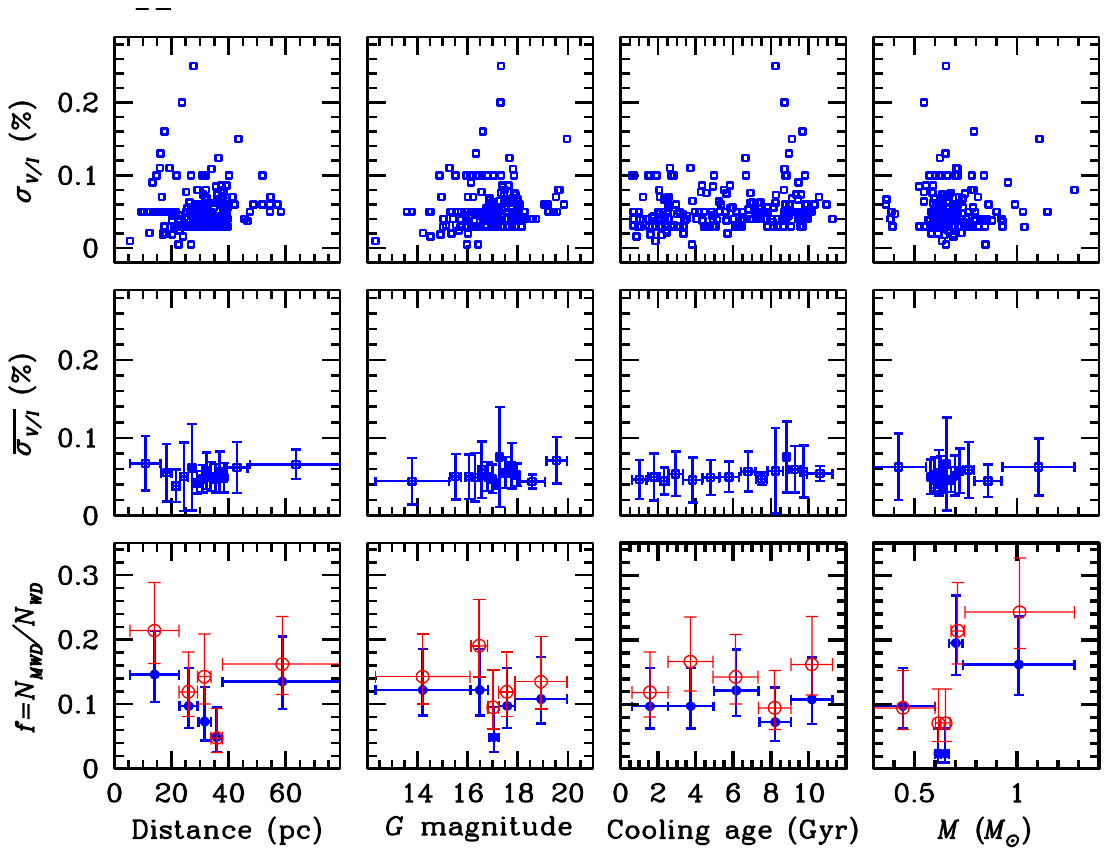}
\end{center}
\caption{\label{Fig_Sigma} Various quantities as a function of distance to the Sun, $G$ magnitude,
cooling age, and stellar mass. 
{\it Top panels:} the smallest uncertainty of any measurement for each individual star.
{\it Middle panels:} typical uncertainties among groups of 15 stars sorted in the order of the abscissa --
the error bars represent the standard deviation of these uncertainties (see text for a more detailed explanation).
{\it Bottom panels:} the inferred frequency $f$ of magnetic stars in bins as defined above, but with size corresponding to about 40 stars;
the vertical error bars represent the 68 per cent Wilson binomial confidence intervals, plotted about the observed fraction $f$; the horizontal bar shows the range of each bin. The two panels highlighted in boldface provide more direct physical information, the remaining ones help to check for potential biases. The top and middle panels include all stars of Table~\ref{Table_Stars}. In the bottom panels, blue symbols represent the results obtained when excluding stars whose polarisation spectra show contributions from spectral lines (marked with \spl\ in Tables~\ref{Table_MWD} and \ref{Table_Stars}) while suspected magnetic white dwarfs are treated as non-magnetic. Red symbols represent the results obtained when including all stars with spectral line contributions and treating suspected magnetic white dwarfs as magnetic.}
\end{figure*}
%%%%%%%%%%%%%%%%%%%%%%%%%%%%%%%%%%%%%%%%%%%%%%%%%%%%%%%%%%%%%%%%%%%%%%%%%%%%%%%%%%%%%%%%%

Table~\ref{Table_Stars} reports stellar parameters and essential information for all the polarimetrically observed DC white dwarfs known to us, together with the maximum observed polarisation value (in a broadband filter), or a zero if no polarisation signal has been detected. We note that more than 90\,\% of DC white dwarfs of Table~\ref{Table_Stars} are within the local 40\,pc volume, and that the same Table includes only 188 of the 287 DC white dwarfs of the local 40\,pc volume. It is important to note that we are not dealing with a volume-limited survey, in particular we think that, owing to the smaller telescopes available, the target list in the Northern Hemisphere is biased against the faintest (hence probably older) white dwarfs. 

Figure~\ref{Fig_Compli} illustrates the distribution of the observed sample relative to the parent population of DC white dwarfs within\,40 pc, and allows us to assess possible age-dependent selection effects. The additional objects observed beyond 40\,pc improve the representation of older stars, but do not contribute to the statistics of the local sample. Figure~\ref{Fig_Compli} suggests that our sample is broadly representative of the local DC population in terms of cooling age distribution. However, the main limitation of this analysis is inherent to the interpretation of the magnetism of DC white dwarfs in the wider context of stellar evolution: as stated elsewhere in this paper, observations of DC white dwarfs, taken in isolation, do not consistently track the evolution of magnetism with age. 

In the absence of detailed models, particularly for continuum circular polarisation, in magnetised cool DC atmospheres, we do not attempt to convert these spectra into quantitative field strengths. We highlight trends in the frequency of magnetism and in the amplitude of the observed polarisation. We recall, however, that the apparent lack of spectral features does not correspond to a specific evolutionary stage; for instance, He-rich stars become featureless at much younger ages than H-rich stars.

As a preliminary step, we investigate whether our polarimetric dataset is affected by systematic observational biases. The top row of Fig.~\ref{Fig_Sigma} shows, for each star, the smallest uncertainty among the available polarimetric measurements. The second row shows the behaviour of the typical uncertainty after binning the data as follows: we sort the stars by increasing value of the quantity of the $x$-axis, we divide them into groups of 15, and compute the mean uncertainty within each group. Finally we plot the mean uncertainty, with its standard deviation shown as the vertical error bar, versus the midpoint of the corresponding $x$ range, with horizontal error bars equal to $(x_{\rm max}-x_{\rm min})/2$. Visual inspection of these plots shows that although our sample is heterogeneous in terms of measurement sensitivities, there is no marked dependence of the typical uncertainty on stellar mass or cooling age.

\subsection{Overall frequency of magnetism in DC white dwarfs}

%%%%%%%%%%%%%%%%%%%%%
\begin{table}
\tabcolsep=0.05cm
\caption{\label{Table_Distance} Frequency of magnetic DC white dwarfs.}
\begin{center}
\begin{tabular}{l|ccr|ccr}
\hline
               &\multicolumn{3}{c}{Sample A}
               &\multicolumn{3}{c}{Sample B} \\
\hline
Distance range & $N_{\rm MWD}$ & $N_{\rm WD}$ &
\multicolumn{1}{c}{$f$} &
$N_{\rm MWD}$ & $N_{\rm WD}$ &
\multicolumn{1}{c}{$f$} \\
\hline
$0 < d \leq 20$ pc
 & 5 & 30 & $16.7\,^{+7.9}_{-5.7}$\,\%
 & 6 & 30 & $20.0\,^{+8.2}_{-6.3}$\,\% \\[2mm]
$20 < d \leq 40$ pc
 & 13 & 155 & $8.4\,^{+2.5}_{-2.0}$\,\%
 & 19 & 158 & $12.0\,^{+2.8}_{-2.4}$\,\% \\[2mm]
$0 < d \leq 40$ pc
 & 18 & 185 & $9.7\,^{+2.4}_{-2.0}$\,\%
 & 25 & 188 & $13.3\,^{+2.7}_{-2.3}$\,\% \\[2mm]
All distances
 & 20 & 201 & $10.0\,^{+2.3}_{-1.9}$\,\%
 & 28 & 205 & $13.7\,^{+2.6}_{-2.2}$\,\% \\
\hline
\end{tabular}
\end{center}

\noindent
$N_{\rm MWD}$ = number of magnetic white dwarfs; $N_{\rm WD}$ = total number of white dwarfs.\\
SAMPLE A: All stars of Table~\ref{Table_Stars} excluding those marked \spl\ (with contributions from spectral lines), and considering the suspected magnetic white dwarfs as non-magnetic. \\
SAMPLE B: All stars of Table~\ref{Table_Stars} including those marked \spl, and considering the suspected magnetic white dwarfs as magnetic ones.
\end{table}
%%%%%%%%%%%%%%%%%%%%%%

We consider two samples: (A) the 201 polarimetrically observed white dwarfs selected on the basis of a previous DC classification, excluding the four objects whose polarisation spectra show substantial contributions from spectral lines, and treating the four suspected magnetic white dwarfs as non-magnetic; (B) the full compilation of 205 stars in Table~\ref{Table_Stars}, including the objects with spectral-line contributions and treating the four suspected magnetic stars as magnetic.

Sample A contains 20 firmly magnetic stars, corresponding to a detected magnetic fraction of $10.0\,^{+2.3}_{-1.9}$\,\%. If the four suspected magnetic stars were confirmed, while retaining the exclusion of the four objects with substantial spectral-line contributions, the fraction would increase to $11.9\,^{+2.5}_{-2.1}$\,\%. Sample B contains 28 magnetic or suspected magnetic stars, corresponding to a fraction of $13.7\,^{+2.6}_{-2.2}$\,\%.

Both limiting estimates are given in Table~\ref{Table_Distance}, where the frequency $f$ is calculated as the number $N_{\rm MWD}$ of magnetic white dwarfs divided by the total number $N_{\rm WD}$ of white dwarfs. The uncertainties represent the 68 per cent Wilson binomial confidence intervals. Depending on the treatment of the suspected magnetic stars and of objects with significant spectral-line contributions, the inferred incidence of detectable magnetism is approximately 10--14\,\%.

\subsection{Trend with mass}
%%%%%%%%%%%%%%%%%%%%%%%%%%%%%%%%%%%%%%%%%%%%%%%%%%%%%%%%%%%%%%%%%%%%%%%%%%%%%%%%%%%%%%%%%
\begin{figure}
\begin{center}
\end{center}
\includegraphics[angle=0,width=8.3cm,trim={1.2cm 6.2cm 1.0cm 3.1cm},clip]{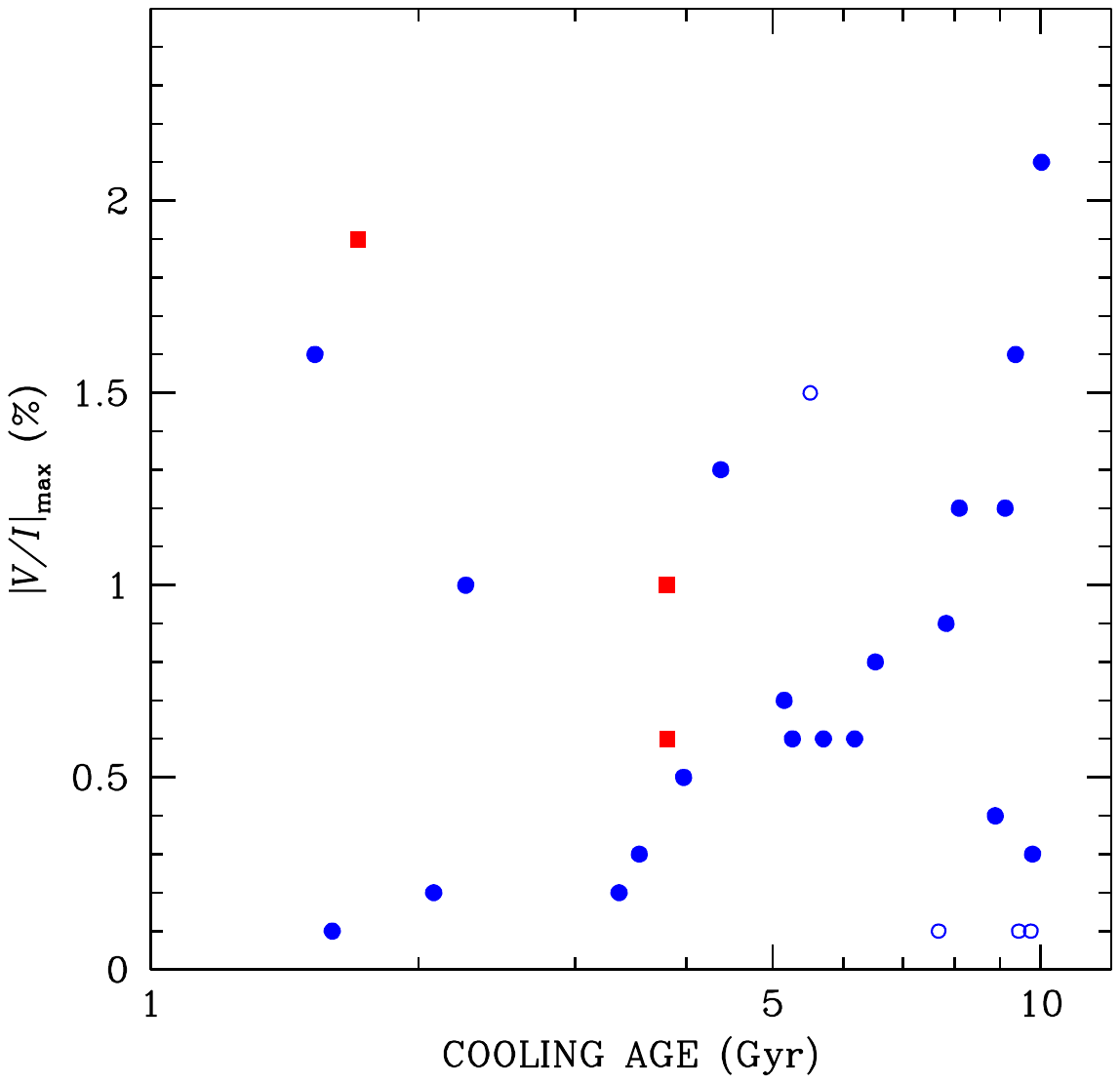}
\caption{\label{Fig_P} Maximum of the absolute value of the broadband polarisation observed in the DC stars of our sample as a function of cooling age.
Blue filled circles represent the genuine DC in which no hint of contribution from lines is seen in the spectra, while the red filled squares represent the stars with some residual spectral features, mainly in the Stokes $V/I$ profiles. Star WD\,0756+437 (which shows residual spectral features) is outside of the plot limits in $\vert V/I\vert_{\rm max}$. Blue empty circles represent stars that are suspected magnetic. Data are from Table~\ref{Table_MWD}.}
\end{figure}
%%%%%%%%%%%%%%%%%%%%%%%%%%%%%%%%%%%%%%%%%%%%%%%%%%%%%%%%%%%%%%%%%%%%%%%%%%%%%%%%%%%%%%%%%
Within our observed sample, the incidence of detectable magnetism is substantially higher among massive white dwarfs than among objects of average or lower mass (see Table~\ref{Table_Mass} and Fig.~\ref{Fig_Sigma}). This is in agreement with a general trend of all white dwarfs outlined in the previous literature \citep[e.g.][]{Lieetal88,OBretal23}. 
%%%%%%%%%%%%%%%%%%%
\begin{table}
\tabcolsep=0.05cm
\caption{\label{Table_Mass} Frequency of magnetic DC
white dwarfs in different mass ranges.}
\begin{center}
\begin{tabular}{l|ccr|ccr}
\hline
               &\multicolumn{3}{c}{Sample A}
               &\multicolumn{3}{c}{Sample B} \\
\hline
Mass range & $N_{\rm MWD}$ & $N_{\rm WD}$ &
\multicolumn{1}{c}{$f$} &
$N_{\rm MWD}$ & $N_{\rm WD}$ &
\multicolumn{1}{c}{$f$} \\
\hline
$M \leq 0.8\,M_\odot$
 & 15 & 182 & $8.2\,^{+2.3}_{-1.8}$\,\%
 & 19 & 183 & $10.4\,^{+2.5}_{-2.0}$\,\%
 \\[2mm]
$M > 0.8\,M_\odot$
 & 5 & 19 & $26.3\,^{+11.1}_{-8.7}$\,\%
 & 9 & 22 & $40.9\,^{+10.7}_{-9.9}$\,\%
 \\[2mm]
$M \leq 0.9\,M_\odot$
 & 18 & 192 & $9.4\,^{+2.3}_{-1.9}$\,\%
 & 23 & 194 & $11.9\,^{+2.5}_{-2.1}$\,\%
 \\[2mm]
$M > 0.9\,M_\odot$
 & 2 & 9 & $22.2\,^{+16.2}_{-10.7}$\,\%
 & 5 & 11 & $45.5\,^{+14.8}_{-14.0}$\,\%
 \\
\hline
\end{tabular}
\end{center}
\end{table}
%%%%%%%%%%%%%%%%%%
\subsection{Trend with cooling age}
Within the statistical uncertainties, the incidence of detectable circular polarisation does not show a monotonic dependence on cooling age (see Fig.~\ref{Fig_Sigma}). Relatively strong circular polarisation signals, with $|V/I|_{\rm max}$ of order 1--2\,\%, are observed not only at intermediate ages but also among stars with cooling ages close to 10\,Gyr, while weak polarisation signals are also found throughout the full age range. Magnetic fields are present even in extremely old white dwarfs. There are now four confirmed magnetic stars with estimated cooling ages above 9 Gyr: 
WD\,0205--283,
WD\,1601--073,
WD\,0042--064, and
WD\,0338-331.
We note that the latter star, WD\,0338--331, has the largest estimated cooling age of any genuine magnetic DC in our sample (10\,Gyr), and it is particularly interesting because it shows one of the largest continuum signals, $\simeq 2.1$\,\%, in the sample.
In conclusion, within our limited sample, there is no evidence that the detectable polarisation signal becomes systematically weaker with cooling age, in fact, one could even argue that Fig.~\ref{Fig_P} suggests that the polarisation amplitude increases with cooling age.

\subsection{Amplitude of the polarisation}
Among the objects in our compilation whose circular polarisation spectra show no identifiable contribution from spectral lines, $|V/I|$ does not exceed approximately 2\,\%. WD\,0756+437 is the only object in the sample showing a much larger amplitude, reaching about 8.5\,\%, but its polarisation spectrum contains prominent line-related features identified with hydrogen transitions in a field of roughly 200–220\,MG. In fact,
circular polarisation amplitudes substantially exceeding about 2\,\% are observed only in white dwarfs whose spectra contain strong spectral features. For example, apart from the case of WD\,0756+437 discussed above, WD\,0316--849 = CL Oct, whose circular polarisation spectrum reaches amplitudes of about 10\% \citep{Feretal97}, is a DA white dwarf; WD\,1900+705 = LAWD 73 = Grw+70$^\circ$\,8247, for which the polarisation reaches about 5\,\%, is also a DA \citep[][and references therein]{BagLan19a}; and WD\,1036--204, whose polarisation exceeds 10\,\% \citep{Lieetal78}, is a DQ white dwarf. 
The existence of an apparent upper limit on the continuum polarisation has two possible explanations that we consider: (1) Field strength may decay with time, and the reason we see lower polarisation in DC white dwarfs is because these are generally older than stars that are still hot enough to show spectral lines. (2) The largest polarisation amplitudes observed in magnetic white dwarfs are associated with spectral line opacity, because this can produce much stronger wavelength-dependent circular polarisation features than magnetic dichroism in the continuum. 

Explanation (1), that the observed ceiling in polarisation reflects a lower field strength in older stars, is not supported by the observations presented here. The maximum continuum polarisation in our sample occurs in one of the oldest DCs, and Fig.~\ref{Fig_P} shows no evidence for a systematic decline of polarisation amplitude with cooling age. We now consider explanation (2), i.e., whether atmospheric physics may provide a natural explanation for the different maximum levels of circular polarisation observed in the coolest MWDs with and without obvious line absorption features. The interpretation of continuum polarisation is difficult because the atmospheric composition of a DC white dwarf may itself be uncertain and, consequently, the continuum opacity sources that determine the observed spectrum are also uncertain \citep[see, for example,][Sect.~2.7]{Kiletal26}. We are therefore far from being able to predict continuum polarisation quantitatively as a function of field strength and geometry. However, the fact that in all the DCs of our sample both the intensity ($I$) and polarisation ($V/I$) spectra vary smoothly with wavelength suggests that the differential opacity experienced by the two polarisation states also varies smoothly across most of the optical spectrum. Such continuum dichroism can produce circular polarisation, but is unlikely to produce by itself the very large local contrast between the opacities of the two states as can occur in a spectral line. Spectral lines provide such a mechanism. At wavelengths corresponding to a strong transition, the line opacity may greatly exceed the local continuum opacity and, in a magnetic field, may also differ strongly between the two polarisation states (this is exactly what is seen in oppositely polarised $\sigma$ components of a Zeeman split line). The resulting contrast can produce very large values of $V/I$, substantially exceeding those generated by the smoother differential opacity of the continuum. This provides a plausible explanation for why polarisation amplitudes above a few per cent are observed predominantly in magnetic white dwarfs with strong spectral features.

\subsection{Variability}\label{Sect_Variability}
More than two thirds of the stars of Table~\ref{Table_Stars} have been observed more than once, either with filter polarimetry or spectropolarimetry. These data allow us to explore questions about polarimetric variability, both on a time scale of the order of $\sim 10$\,min and on a time scale of the order of days, weeks, or even years. 
In Table~\ref{Table_MWD} we report the clearest findings, i.e., stars that show no hint of variability (label "nv") and stars that show signs of variability (labels  'v' and 'v:'). \citet{BagLan24} have found that many old, strongly magnetic white dwarfs have  magnetic fields that are symmetric about the rotation axis. They proposed that magnetic field stresses distort the star and define a preferred principal axis of inertia. If the rotation axis is initially misaligned with this axis, internal dissipation damps the resulting free precession and may drive the star towards rotation about a principal inertia axis, or perhaps more likely, drive the magnetic field to be symmetric about the rotation axis. 

Most average-mass magnetic stars for which adequate repeated observations are available show no evidence of polarimetric variability, although the relatively young WD\,0912+536 and WD\,1136--129 are clear exceptions. Several of the more massive magnetic objects show definite or suspected variability. A firmer assessment of the variability of these (and other) white dwarfs requires polarimetric time series with short-exposure frames, which are the subject of future work.

\subsubsection{Use of non-variable magnetic DC white dwarfs as standard stars for circular polarisation} 
While several catalogues of standard stars exist for linear polarisation \citep[e.g.,][]{HsuBre82,Witetal92,Fosetal07,Ciketal17,Cotetal24}, no equivalent catalogue is available for circular polarisation. In principle, Table~\ref{Table_MWD} could be used to identify candidate circular-polarisation standards among magnetic white dwarfs. However, a caveat should be kept in mind when comparing filter-polarimetric measurements obtained at different epochs. 
In this observing mode, the measured circular polarisation is a bandpass-integrated quantity,
%%%%%%%%%%%%%%%%%%%%%%%%%%%%%%%%%%%%%%%%%%%%%%%%
\begin{equation}
\left\langle\frac{V}{I} \right\rangle_T =
\frac{\int {\rm d}\lambda\ \frac{V(\lambda)}{I(\lambda) }\, I(\lambda)\, T(\lambda)}
{\int {\rm d}\lambda\ I(\lambda) \, T(\lambda)}
\label{Eq_FilterPol}
\end{equation}
%%%%%%%%%%%%%%%%%%%%%%%%%%%%%%%%%%%%%%%%%%%%%%%%
where $T(\lambda)$ is the effective transmission function, including the atmosphere, telescope and instrument optics, filter transmission, detector response, and any time-dependent changes in these components.  In magnetic white dwarfs, $V(\lambda)/I(\lambda)$ may rapidly change with wavelength, therefore a change in $T(\lambda)$ can affect the measured band-integrated polarisation. This effect may occur not only when comparing data obtained with different instruments, but also for observations obtained with the same instrument and nominal setup, if the effective transmission varies with time or with observing conditions, for instance because of different airmasses.
Some apparent long-term variability in filter-polarimetric data could therefore be caused by changes in the effective passband rather than by intrinsic variability of the star. For this reason, comparisons based on filter polarimetry should be restricted, as far as possible, to data obtained with the same instrument and setup, and should still be interpreted with caution. Robust tests of temporal constancy are best based on spectropolarimetry, where the comparison is made within narrow wavelength bins and the effective transmission can be assumed to be nearly constant across each bin.

\section{Conclusions}
We have presented a new circular-polarimetric survey of cool, featureless DC white dwarfs, consisting of 130 FORS2 observations (155 polarimetric measurements) of 97 stars. The observations were obtained either in filter-polarimetric mode or in low-resolution spectropolarimetric mode. Most of the targets had never previously been observed in polarimetry. We confirmed circular polarisation in 11 stars that were already known to be magnetic from previous work, and discovered seven new magnetic DC white dwarfs. Two additional objects of our survey show a weak polarisation signal that requires confirmation.

By combining our new measurements with circular polarisation observations from the literature, we analysed a sample of 205 white dwarfs previously classified as DC. Among these, 24 show firmly detectable circular polarisation, and another four are suspected magnetic. The inferred magnetic incidence depends on the adopted sample definition, as some apparently featureless stars reveal polarised spectral lines, while the magnetic nature of four objects remains uncertain. Depending on the treatment of these objects, we find that approximately 10--14\,\% of the DC-selected white dwarfs in our sample exhibit detectable magnetic fields. Although our work does not constitute a complete volume-limited survey, our sample appears broadly representative of the local DC population, at least in terms of cooling-age distribution. We therefore suggest that a similar incidence of detectable magnetism may characterise the underlying population of DC white dwarfs.

Using only the data presented in this work, we cannot accurately investigate whether the incidence of magnetism changes with cooling age. However, our data do not suggest a decline in the occurrence of detectable magnetism among the oldest objects in the sample. Four confirmed magnetic objects in our compilation have estimated cooling ages above 9\,Gyr, and the oldest, WD\,0338--331, has an estimated cooling age of 10\,Gyr while exhibiting continuum circular polarisation reaching approximately 2\,\%.  Magnetic DC white dwarfs are thus present across the full age range probed here. Similarly, the observed polarisation amplitudes do not show an obvious decrease towards old ages. Relatively strong signals, with $|V/I|_{\rm max}$ of order $1-2$\,\%, are found among very old stars as well as at intermediate ages. In summary, our results show that strong, detectable magnetic fields can persist to the latest observable stages of white dwarf cooling, although they do not provide a quantitative constraint on magnetic-field decay.

In contrast, the incidence of detectable magnetism does appear to depend on stellar mass. The magnetic fraction is lower among low-mass and average-mass DC white dwarfs, and higher among stars more massive than the canonical white dwarf mass. The highest values of $|V/I|_{\rm max}$ are also found mainly among stars more massive than average. 

An important empirical result is that no object whose polarisation spectrum appears to be dominated by genuine continuum polarisation shows $|V/I|$ substantially above 2\,\%; the only much more strongly polarised object in the sample, WD\,0756+437, displays prominent line-related polarisation structure.

A limitation of the present work is that continuum circular polarisation cannot be converted into magnetic field strength in a model-independent way. Moreover, the available measurements are heterogeneous, including broadband filter polarimetry and spectropolarimetry obtained with different setups. For this reason, our results should be interpreted as constraints on the incidence and amplitude of detectable circular polarisation in featureless stars, rather than as a census of magnetic-field strengths. Detailed models of continuum circular polarisation in cool magnetic atmospheres will be needed to turn these empirical trends into more quantitative constraints on the origin and long-term evolution of magnetic fields in white dwarfs. Despite these limitations, the combined dataset provides the most complete polarimetric view so far of magnetism among cool DC white dwarfs.

\section*{Acknowledgements}
The new observations presented in this work were made with the FORS2 instrument at the ESO Telescopes at the La Silla Paranal Observatory under program ID
109.235Q.001, 110.243J.001, 110.23XV.001, 110.23XV.002, 111.24PQ.001, 111.24PQ.002, 111.24G5.001, 111.24G5.002, 112.25C9.001, 114.2713.001, 114.2713.003, 115.27UN.002, 115.27UN.003 (PI=Bagnulo); 0103.D-0029(B), 0104.D-0298(A), 0104.D-0298(E) (PI=Landstreet); and  114.27BU.001 (PI=Schreiber). This research has also made use of additional FORS2 data obtained from the ESO Science Archive Facility, program ID=080.D-0521(A) (PI=Kawka). JDL acknowledges past financial support of the Natural Sciences and Engineering Research Council of Canada (NSERC), most recently with funding reference number 6377-2016. 

%%%%%%%%%%%%%%%%%%%%%%%%%%%%%%%%%%%%%%%%%%%%%%%%%%
\section*{Data Availability}
The raw FORS2 observations and associated calibration files are available from the ESO Science Archive Facility (archive.eso.org). References to previously published observations are provided in Table~\ref{Table_Stars}.

\bibliographystyle{mnras}
\bibliography{sbabib} 
%%%%%%%%%%%%%%%%%%%%%%%%%%%%%%%%%%%%%%%%%%%%%%%%%%

\newpage 

%%%%%%%%%%%%%%%%% APPENDICES %%%%%%%%%%%%%%%%%%%%%
\appendix
\section{Observing log}
Table~\ref{Table_WD0912} shows the log of the observations of WD\,0912+536 obtained with Palomar MCSP in the 1970s.
%%%%%%%%%%%%%%%%%%%%%%%
\begin{table}
  \caption{\label{Table_WD0912} Circular polarised spectra of WD\,0912+536 = G195-19
    obtained with Palomar MCSP by Angel \& Landstreet in the 1970s. Phases computed using
    JD0 = 2440979.071 as given by \citet{Angetal72-eph} and the period $P =$ 1.331045(3) from \citet{Faretal24}.
    The third column is the duration of each observation; the exact exposure times are not available.}
\begin{center}
  \begin{tabular}{ccrc}
\hline\hline
Date & UT& Duration & Rotation      \\
     &   &          & phase         \\
\hline
1972-02-07 & 11:12 & 249\,m & 0.406 \\
1973-11-06 & 11:40 & 171\,m & 0.744 \\
1973-11-07 & 11:35 & 198\,m & 0.493 \\
1973-11-08 & 12:48 &  64\,m & 0.281 \\
1975-04-30 & 05:30 & 157\,m & 0.247 \\
1975-05-01 & 05:43 & 182\,m & 0.005 \\
\hline
\end{tabular}
\end{center}
\end{table}
%%%%%%%%%%
\begin{table}
\caption{\label{Table_Grisms} Characteristics of the FORS1/2 instrument settings (300V, 300I, 1200B, 600B grisms) and of the ISIS setting (R600B and R1200R gratings). }
\begin{center}
\begin{tabular}{llcc}
\hline
\hline
Instrument & Grism/   &    $R$    & Spectral    \\
           & Grating  &(1\arcsec) & range (\AA) \\
\hline
FORS2 & 300V     &  440      & 3800--9200 \\
FORS2 & 300I     &  660      & 6000--11000 \\
FORS2 & 1200B    & 1420      & 3700--5100 \\
FORS1 & 600B     &  780      & 3600--5200 \\
FORS2 & 600B     &  780      & 3600--6200 \\
ISIS  & R600B    & 2600      & 3500--5200 \\
ISIS  & R1200R   & 8600      & 5800--6200 \\
\hline
\end{tabular}
\end{center}
\end{table}
%%%%%%%%%%%%%%%%%%%%%%%%%%%%%%%%%%%%%%%

In Table~\ref{Table_Log} we report the log of all FORS observations of DC white dwarfs published here for the first time, either in imaging polarimetry or spectropolarimetry mode. For convenience, we report also all FORS and ISIS spectropolarimetry of DC white dwarfs previously published. Table~\ref{Table_Grisms} summarises the characteristics of the instrument settings. In Table~\ref{Table_Grisms}, the spectral resolving power $R$ refers to a 1\arcsec\ slit width, but most of the observations with FORS were obtained with a 2\arcsec slit width (polarisation spectra of DC white dwarfs are intrinsically very smooth). 

All FORS spectropolarimetric measurements were obtained with the MIT CCD, except for those of WD\,0004+122 with the 600B grism on 2024-10-28 and that with the 300V grism on 2024-10-30, and all those obtained by \citet{KawVen12} in 2007 and 2008, that were obtained with the EEV CCD, optimised for the blue. Data obtained with the EEV CCD beyond 6500\,\AA\ are affected by strong fringing and should be considered with caution. 

All filter polarimetry observations were made with the MIT CCD (optimised for the red), and all filter polarimetry observations were obtained with readout mode 1x1, except for 
WD\,0042--064,
WD\,0205--283 on 2025-01-27,
WD\,0338--331 on 2025-01-03,
WD\,1218+095,
WD\,1825--607,
WD\,1946--735, 
that were observed with the standard 2x2 readout mode. 

WD\,2156$-$028 and WD\,2156$-$029 are close enough to be imaged in the FORS2 field of view hence their observations were obtained simultaneously.

Results obtained in spectropolarimetric mode are reported as filter integrated measurements calculated using Eq.~(\ref{Eq_FilterPol}) and adopting the transmission functions of filter b\_high, v\_high, R\_SPECIAL and I\_BESS as given in the ESO FORS instrument webpages. The corresponding errors are dominated by non-photon-noise effects, particularly cosmic rays and instrumental polarisation introduced by the optics preceding the polarimetric optics. A reasonable estimate of these errors is about 0.05\,\% \citep[instrumental circular polarisation was discussed in Sect.~3 of][]{Sieetal14} in the v, R, and I filters, although potential effects of cosmic rays may be difficult to quantify. Uncertainties tend to be higher (0.1--0.2\,\%) at wavelengths corresponding to the b\_high filter, where instrumental polarisation may be higher. Furthermore, at wavelengths shorter than 4200\,\AA, the transmission of the MIT CCD drops considerably (see Fig.~2.8 of the instrument manual). Detections are highlighted by the use of boldface fonts for the target's name. The detections of WD\,1315+222 and WD\,2049--222 rely primarily on broadband circular polarisation measurements reported in previous studies. 
%%%%%%%%%%%%%%%%
\onecolumn
\begin{center}
\tabcolsep=0.05cm
\begin{longtable}{llccr
r@{}c@{}l
r@{}c@{}l
r@{}c@{}l
r@{}c@{}l
l}
\caption{\label{Table_Log} Polarimetric measurements of DC WDs (BBCP stands for Broadband Circular Polarisation). Names in boldface refer to stars in which a polarisation signal was firmly detected, see text for more details.}\\
\hline\hline
STAR                             &
Grism/                           &
DATE                             &
UT                               &
Exp                              &
\multicolumn{12}{c}{BBCP (\%)}   &
REF.                             \\
                                 &
Filter                           &
yyyy-mm-dd                       &
hh:mm                            &
(s)                              &
\multicolumn{3}{c}{b}            &
\multicolumn{3}{c}{v}            &
\multicolumn{3}{c}{R}            &
\multicolumn{3}{c}{I}            &
                                 \\
\hline
\endfirsthead
\caption{continued.}\\
\hline\hline
STAR                             &
Grism/                           &
DATE                             &
UT                               &
Exp                              &
\multicolumn{12}{c}{BBCP}        &
REF.                             \\
                                 &
Filter                           &
yyyy-mm-dd                       &
hh:mm                            &
(s)                              &
\multicolumn{3}{c}{B}            &
\multicolumn{3}{c}{V}            &
\multicolumn{3}{c}{R}            &
\multicolumn{3}{c}{I}            &
                                 \\
\hline
\endhead
\hline
\endfoot
%        1         2         3         4         5         6       8         9        10

%                                                               B                 V                 R                  I
%

%-----------------------------------------------------------------------------------------------------------------%-
{\bf WD\,0004$+$122}& 300V  & 2019-10-07 & 01:17 & 1560 &\va{ 0.50}{0.12} &\va{ 0.51}{0.10} &\va{ 0.78}{0.10} &\va{-0.23}{0.10} & BL20 \\
                    & R     & 2023-11-11 & 00:17 &  240 &\nulla           &\nulla           &\vi{ 0.87}{0.04} &\nulla           & \unp \\ % 7.0     0.863    0.033    0.026    0.026 M
                    & 300V  & 2024-10-04 & 02:16 & 1600 &\va{ 0.71}{0.10} &\va{ 0.80}{0.10} &\va{ 0.85}{0.10} &\va{-0.22}{0.10} & \unp \\
                    & 300I  & 2024-10-04 & 02:41 &  720 &\nulla           &\nulla           &\va{ 0.86}{0.10} &\va{-0.20}{0.10} & \unp \\
                    & 600B  & 2024-10-28 & 00:00 & 1200 &\va{ 0.69}{0.10} &\va{ 0.73}{0.10} &\nulla           &\nulla           & \unp \\
                    & 300V  & 2024-10-30 & 00:16 & 1200 &\va{ 0.76}{0.10} &\va{ 0.83}{0.10} &\va{ 0.90}{0.10} &\va{-0.16}{0.10} & \unp \\ 
%----------------------------------------------------------------------------------------------------------------------------------------
{\rm WD\,0024$+$056}& R     & 2023-10-20 & 02:04 & 1320 &\nulla           &\nulla           &\vi{ 0.01}{0.03} &\nulla           & \unp \\ % 8.0     0.001    0.033    0.039   -0.039 
%------------------------------------------------------------------------------------------
{\rm WD\,0034$-$285}& R     & 2023-10-08 & 03:45 &  920 &\nulla           &\nulla           &\vi{ 0.00}{0.04} &\nulla           & \unp \\
\end{longtable}
%%%%%%%%%%%%%%%%%%%%%%%%%%%%%%%%%%%%%%%%%%%%%%%%%%%%%%%%%%%%%%%%%%%%%%%%%%%%%%%%%%%%%%%%%%%%%%%%%%%%%%%%%%%%%%
\end{center}
   \begingroup
    \let\clearpage\relax 
    \onecolumn 
{\bf Keys to references:}
\tw: this work;
BL18: \citet{BagLan18};
BL20: \citet{BagLan20};
BL21: \citet{BagLan21};
BL24: \citet{BagLan24};
KV12: \citet{KawVen12}.
     \endgroup
    
\newpage

\section{Summary Table}
\begin{small}
\begin{center}  % 1234567890
\begin{longtable}{llrrrrrcl}
\caption{\label{Table_Stars} Atmospheric parameters for DC white dwarfs observed in circular polarisation, together with the maximum absolute value of the polarisation and a reference to the observations. This compilation does not represent a volume-limited sample, and target selection is at least partially biased against the faintest stars of the Northern Hemisphere -- see text for details.}\\
\hline\hline
STAR                             & %1
Main Simbad                      & %2
$G$                              & %4
$d$                              & %5
$T_{\rm eff}$                     & %6
$M$                              & %7
Age                              & %8
$\vert V/I \vert_{\rm max}$       & %9
REF.                             \\ %10
                                 & %Star
Identifier                       & %ident.
                                 & % mag
(pc)                             & %distance
(K)                              &
($M_\odot$)                      &
(Gyr)                            &
(\%)                             &
                                 \\
\hline
\endfirsthead

\caption{continued.}\\
\hline\hline
STAR                             &
Main Simbad                      &
$G$                              &
$d$                              &
$T_{\rm eff}$                     &
$M$                              &
Age                              &
$\vert V/I \vert_{\rm max}$       &
REF.                             \\
                                 & %Star
Identifier                       & %ident.
                                 & % mag
(pc)                             & %distance
(K)                              &
($M_\odot$)                       &
(Gyr)                            &
(\%)                             &
                                 \\
\hline
\endhead
\hline
\endfoot
%        1         2                                   3        4        5       6       7       8      9
WD0000-345   &LAWD 1                                &14.88    &14.8   & 6332  &0.662  & 2.86  &   0    &Sch+01              \\%  2312821266302222720
WD0004+122   &LP 464-57  \splc                      &16.25    &17.5   & 5285  &0.813  & 7.83  & 0.9    &BL20: 300V; \tw: R,600B,300Vx2,300I\\%  2766234439302571904
WD0004+337   &LP 240-29                             &17.23    &34.0   & 5666  &0.705  & 5.26  & 0.6    &Be+24               \\%  2875903332533220992
WD0005+395   &LP 240-30                             &16.56    &34.4   & 5106  &0.287  & 1.78  &   0    &Be+24               \\%   383108338321272448
WD0010+543   &LSR J0013+5437                        &17.99    &32.3   & 4348  &0.631  & 9.44  &   0    &Be+24               \\%   420531621029108608
WD0019+264   &LP 349-1                              &17.46    &32.9   & 5101  &0.660  & 6.90  &   0    &Be+22               \\%  2855790657816672000
WD0019+423   &EGGR 459                              &16.37    &34.5   & 5665  &0.405  & 2.08  &   0    &Ptn97; Be+24$^1$    \\%   385105360675267840
WD0021+683   &G 242-54B                             &17.46    &36.0   & 5663  &0.747  & 5.65  &   0    &Be+24$^1$           \\
\hline
\end{longtable}
%%%%%%%%%%%%%%%%%%%%%%%%%%%%%%%%%%%%%%%%%%%%%%%%%%%%%%%%%%%%%%%%%%%%%%%%%%%%%%%%%%%%%%%%%%%%%%%%%%%%%%%%%%%%%%
\end{center}
\end{small}
    \begingroup
    \let\clearpage\relax 
    \onecolumn 
1 WD\,0019+423, WD\,0021+683, and WD\,0052+595 were previously declared magnetic or suspected magnetic by \citet{Beretal24}, but Stroet et al. (in prep.) have found that, in fact, no polarisation signal could be firmly detected.\\
{\bf Keys to references:}
\tw: this work;
AL71a: \citet{AngLan71-Second};
AL71b: \citet{AngLan71-Periodic};
Ang+70: \citet{AngLan70Further};
Ang+72: \citet{Angetal72-eph};
Ang+81: \citet{Angetal81};
BL18: \citet{BagLan18};
BL20: \citet{BagLan20};
BL21: \citet{BagLan21};
BL24: \citet{BagLan24};
Be+22: \citet{Beretal22};
Be+23: \citet{Beretal23};
Be+24: \citet{Beretal24};
KV12: \citet{KawVen12};
LS80: \citet{LieSto80};
Ptn95: \citet{Putney95};
Ptn97: \citet{Putney97};
Sch+01: \citet{Schetal01}.\\
The instrument settings of the observations  presented in this work (marked \tw) and those by \citet{BagLan18,BagLan20,BagLan21,BagLan24} are explicitly specified in the Reference column. 
For FORS filter polarimetry we use the following keys: R for R\_Special, v for v\_high, b for b\_high and I for I\_BESS filter described in the instrument manual; 
for spectropolarimetry with FORS we refer to the grism names 300V, 300I, 600B, 1200B -- slit width was generally 1\arcsec\ with 1200B and 600B, and 2\arcsec\ with 300V and 300I; 
the label IS refers to the spectra obtained with the ISIS instrument and the R600B and R1200R gratings simultaneously, and the slit width was generally set to 1\arcsec.\\
All observations obtained by \citet{Beretal22,Beretal23,Beretal24} were obtained simultaneously in the $B^\prime V^\prime R^\prime$ filters described in \citet{Beretal22}.\\
\citet{AngLan71-Second},
\citet{Putney97} and 
\citet{Schetal01} observed in spectropolarimetric mode. \\
\citet{LieSto80} used filter polarimetry with an optical white filter;
\citet{AngLan71-Periodic}, \citet{Angetal72-eph} and \citet{Angetal81} used filter polarimetry.\\
A ``0'' in the $\vert V/I \vert_{\rm max}$ column means that no polarisation signal was detected within $3\,\sigma$; for the sensitivity limit we refer to the original reference.\\
     \endgroup

%%%%%%%%%%%%%%%%%%%%%%%%%%%%%%%%%%%%%%%%%%%%%%%%%%
% Don't change these lines
\bsp	% typesetting comment
\label{lastpage}
\end{document}